\documentclass{SciPost}

\newcommand{\linenumbermode}{\linenumbers}
\renewcommand{\linenumbermode}{\nolinenumbers}  

\hypersetup{
    colorlinks,
    linkcolor={red!50!black},
    citecolor={blue!50!black},
    urlcolor={blue!80!black}
}

\usepackage{amssymb}
\usepackage{bm}
\usepackage{booktabs}
\usepackage{fontawesome5}
\usepackage{siunitx}
\usepackage{listings}
\usepackage[dvipsnames]{xcolor}

\DeclareSIUnit\eV{e\kern-.05em V}

\fancypagestyle{SPstyle}{
\fancyhf{}
\lhead{\colorbox{scipostblue}{\bf \color{white} ~SciPost Phys. Codebases 90 (2026) }}
\rhead{}

\fancyfoot[C]{\textbf{\thepage}}
}

\graphicspath{{./}}

\newcommand{\notebooklink}[1]{\href{https://github.com/bakerem/spectroxide/tree/v1.0.0/#1}{\faGithub}}

\newcommand{\dc}{\text{DC}}
\newcommand{\br}{\text{BR}}
\newcommand{\Te}{T_\mathrm{e}}
\newcommand{\Tz}{T_\gamma}
\newcommand{\te}{\theta_\mathrm{e}}
\newcommand{\tz}{\theta_\gamma}
\newcommand{\npl}{n_\mathrm{pl}}
\newcommand{\dn}{\Delta n}
\newcommand{\drho}{\Delta\rho/\rho}
\newcommand{\re}{\rho_\mathrm{e}}
\newcommand{\xe}{x_\mathrm{e}}
\newcommand{\Gbb}{G_\mathrm{bb}}
\newcommand{\Gth}{G_\mathrm{th}}
\newcommand{\Jmu}{J_\mu}
\newcommand{\Jbb}{J_\mathrm{bb}^*}
\newcommand{\Jy}{J_y}
\newcommand{\Ysz}{Y_\mathrm{SZ}}
\newcommand{\kc}{\kappa_\mathrm{c}}
\newcommand{\bmu}{\beta_\mu}

\newcommand{\Gpl}[1]{G_{#1}^{(0)}}
\newcommand{\Ipl}[1]{I_{#1}^{(0)}}
\newcommand{\cosmotherm}{\textsc{CosmoTherm}}
\newcommand{\spectroxide}{\textsc{spectroxide}}
\newcommand{\code}[1]{\texttt{#1}}

\begin{document}

\pagestyle{SPstyle}

\begin{center}{\Large \textbf{\color{scipostdeepblue}{
\textsc{spectroxide}: A code package for computing cosmic microwave background spectral distortions
}}}\end{center}

\begin{center}\textbf{
Ethan~Baker\textsuperscript{1$\star$},
Hongwan~Liu\textsuperscript{1$\dagger$}, and
Siddharth~Mishra-Sharma\textsuperscript{2,1$\ddagger$}
}\end{center}

\begin{center}
{\bf 1} Physics Department, Boston University, Boston, MA 02215, USA
\\
{\bf 2} Faculty of Computing \& Data Sciences, Boston University, Boston, MA 02215, USA
\\[\baselineskip]
$\star$ \href{mailto:ebaker@bu.edu}{\small ebaker@bu.edu}
$\dagger$ \href{mailto:hongwan@bu.edu}{\small hongwan@bu.edu}
$\ddagger$ \href{mailto:smishras@bu.edu}{\small smishras@bu.edu}
\end{center}

\section*{\color{scipostdeepblue}{Abstract}}
\textbf{\boldmath{%
We present \spectroxide{}, a code package for computing cosmic microwave background spectral distortions in which all ${\sim}12{,}000$ lines of Rust code, Python interface, and over 950 automated tests were written by an AI assistant (Claude Code) under human physicist supervision.
The solver evolves the photon Boltzmann equation under Compton scattering, double Compton emission, and Bremsstrahlung from $z \sim 5 \times 10^6$ to the present, computing spectral distortions from arbitrary heat and photon injection within this redshift range.
No fully open-source code of this kind is publicly available; we validate against analytic limits, published spectra, and publicly available precomputed Green's function tables.
We document the development as a case study in AI-assisted scientific computing, highlighting how domain expertise caught physics bugs (incorrect dimensional prefactors, near-cancellation errors) that evaded the full automated test suite, and document the practices and validation methodology we found effective for human--AI collaborative development of scientific software.
We make \textsc{spectroxide} publicly available on GitHub.~\href{https://github.com/bakerem/spectroxide}{\faGithub}
}}

\vspace{\baselineskip}

\noindent\textcolor{white!90!black}{%
\fbox{\parbox{0.975\linewidth}{%
\textcolor{white!40!black}{\begin{tabular}{lr}%
  \begin{minipage}{0.54\textwidth}%
    {\small Copyright E. Baker \textit{et al}. \newline
    This work is licensed under the Creative Commons \newline
    \href{http://creativecommons.org/licenses/by/4.0/}{Attribution 4.0 International License}. \newline
    Published by the SciPost Foundation.}
  \end{minipage} & \begin{minipage}{0.40\textwidth}
    {\small Received 2026-05-08 \newline Accepted 2026-11-02 \newline Published 2026-12-02 \newline \href{https://doi.org/10.21468/SciPostPhysCodeb.90}{doi:10.21468/SciPostPhysCodeb.90}}%
  \end{minipage}
\end{tabular}}
}}
}

\linenumbermode

\vspace{10pt}
\noindent\rule{\textwidth}{1pt}
\tableofcontents
\noindent\rule{\textwidth}{1pt}
\vspace{10pt}

\section{Introduction}
\label{sec:intro}
The cosmic microwave background (CMB) is the most precisely measured blackbody in nature: the COBE/FIRAS instrument confirmed its Planckian spectrum to better than \qty{50}{ppm}~\cite{Fixsen1996,Fixsen2009}.
Yet any energy exchange between the photon field and its surroundings---whether from known Standard Model processes or exotic new physics---must alter this spectrum.
If this energy injection occurs sufficiently early ($z \gtrsim 2 \times 10^6$), number-changing processes efficiently restore a Planck spectrum at a slightly different temperature.
Energy injection after this epoch instead imprints slight deviations from the Planck spectrum.
These \emph{spectral distortions} encode the thermal history of the universe from $z \sim 2 \times 10^6$ to the present, providing information complementary to CMB anisotropies and large-scale structure~\cite{Chluba2014review,Chluba2021Voyage}.

Within $\Lambda$CDM, Silk damping of primordial acoustic waves produces a guaranteed $\mu$-distortion at the $\mu \sim 2 \times 10^{-8}$ level~\cite{Sunyaev1970mu,Zeldovich1969,Silk:1968,Sunyaev1970dissipation,Daly1991,Hu1993,Chluba2012dissipation,Chluba2012cosmotherm}, with a smaller negative contribution $\mu \sim -3 \times 10^{-9}$ from adiabatic cooling of matter relative to radiation~\cite{Chluba:2016bvg}.
At lower redshifts, the thermal Sunyaev--Zel'dovich effect from reionization and structure formation produces an integrated $y$-distortion of order $y \sim 10^{-6}$~\cite{Hill2015}.
Beyond $\Lambda$CDM, spectral distortions can probe dark matter annihilation and decay~\cite{Ellis:1990nb,McDonald2001,Chluba2013DM,Acharya2019,Acharya2019decay,Liu2023exotic}, dark photon oscillations~\cite{Mirizzi2009,Caputo2020,Caputo2020modeling,Chluba2024darkphoton,Arsenadze2025}, axion-like particle conversions~\cite{Tashiro2013axion,Mukherjee2018,Cyr2024axion}, photon injection from particle decays~\cite{Chluba2015photon,Bolliet2021}, evaporating primordial black holes~\cite{Acharya2020pbh}, cosmic strings and other topological defects~\cite{Cyr2023strings}, primordial magnetic fields~\cite{Jedamzik2000}, and first-order cosmological phase transitions~\cite{AminGrin2014}---physics that is in many cases difficult to access by other means (see Refs.~\cite{Chluba2014review,Chluba2021Voyage} for reviews).

A new generation of spectral distortion experiments could more fully realize the scientific potential outlined above.
COBE/FIRAS established the CMB as a blackbody to $|\Delta I/I| \lesssim 5 \times 10^{-5}$~\cite{Fixsen1996}, but the predicted $\Lambda$CDM signals lie below this threshold.
Several concepts have been developed that have sensitivities significantly beyond that of FIRAS, including the space-based PIXIE~\cite{Kogut2011}, the balloon-borne pathfinder BISOU~\cite{BISOU2021}, and SPECTER~\cite{SPECTER2024}.
Most recently, FOSSIL~\cite{FOSSIL2024,Aghanim:2026mii}, a new proposal to ESA's M8 call, targets a sensitivity to $\mu$-type distortions hundreds of times greater than FIRAS.
This would be sufficient to detect the guaranteed Silk damping signal, and would provide a novel probe of BSM physics in the early universe from $z \sim 2 \times 10^6$ to the present day~\cite{Chluba2021Voyage}.

Computing spectral distortions requires solving the photon Boltzmann equation with Compton, double Compton (DC), and Bremsstrahlung (BR) processes.
This is numerically challenging: given the COBE/FIRAS limits, the physical distortion signal is limited to be $\sim 10^5$ times smaller than the Planck background, and DC/BR emission rates diverge as $1/x^3$ at low frequencies, introducing extreme stiffness.
The tremendous progress in spectral distortion science over recent years has been built primarily on the code \cosmotherm{}~\cite{Chluba2012cosmotherm}.\footnote{Interested users should visit \url{www.Chluba.de/CosmoTherm} for more information about accessing \cosmotherm{}.} Green's functions for heat injection processes obtained from this code are publicly available and have been integrated into the Boltzmann code \textsc{class}, following the work of Refs.~\cite{Chluba2013greens,Lucca2020}. However, despite the underlying physics being well-understood and partial numerical results being publicly available, no open-source code currently solves the full Boltzmann equation for spectral distortions. Such a code would be valuable for any work exploring new-physics effects beyond heat injection. A clear example is photon injection, for which no public Green's functions exist, although Ref.~\cite{Chluba2015photon} studied them and provides partial results in plots.

At the same time, the rapid growth in the capabilities of large language models (LLMs) has opened up new possibilities for scientific computing.
Building a PDE solver for spectral distortions would have previously required significant human coding effort; with the current generation of LLMs, however, it is now possible for LLMs to write the vast majority of such a code, simply based on the published literature. 
Spectral distortions are a unique test case for LLM-assisted scientific computing: there is an extensive and mature literature on the underlying physics, but there is no open-source numerical implementation of the underlying physics. 
A spectral distortion solver would therefore have little existing codebase to reference and validate against, but a large body of published physics and partial results that a human expert can use to test the code. 

With this motivation in mind, we present \spectroxide{}, an open-source spectral distortion solver, written in Rust with zero production dependencies.
The code handles heat and photon injection, and is validated against analytic limits, Green's function approximations~\cite{Chluba2013greens,Chluba2015photon,Arsenadze2025}, and \cosmotherm{} public results where available across all thermalization regimes (Sec.~\ref{sec:validation}).
The project was developed entirely through a human--AI collaboration using Claude Code~\cite{ClaudeCode} (Anthropic's command-line coding agent, powered by Claude Opus 4.6 during initial development and 4.7 for subsequent refinement), with the AI handling implementation, testing, debugging, and paper drafting under human physics supervision.
We chose Rust for its compiled performance (comparable to C/Fortran), memory safety guarantees, and built-in test framework, noting that the choice of a compiled language over Python or JAX is less consequential when an LLM mediates the development: contributors need not be fluent in Rust to extend the code.
Users can interact with \spectroxide{} at two levels: the Python interface and tutorial notebooks (Appendix~\ref{app:python}) allow direct use for science without any AI tooling, while contributors adding new physics are encouraged to use an LLM with the project's curated context file (\code{CONTRIBUTING\_CLAUDE.md}), which encodes the numerical pitfalls and testing philosophy needed to produce correct contributions (Sec.~\ref{sec:ai_contributing}).
Recent work has demonstrated that LLMs can re-implement existing solvers when the original source code is available as a reference~\cite{clax2025}.
\spectroxide{} represents a distinct and more challenging case: the solver was built entirely from the published literature, with no full software reference implementation to validate against.
The lessons learned---particularly about the failure modes of AI-generated scientific code and the currently irreplaceable role of domain expertise in catching subtle physics errors that evade extensive automated testing---are documented in detail as one of the first case studies in AI-assisted development of scientific software (Sec.~\ref{sec:ai}).

The paper is organized as follows.
Section~\ref{sec:physics} describes the physics: the photon Boltzmann equation with Compton, DC, and BR scattering.
Sections~\ref{sec:heat_injection} and~\ref{sec:photon_injection} describe the two injection mechanisms implemented in \spectroxide{}---heat injection and photon injection---including how each enters the PDE solver and the specific scenarios supported.
Section~\ref{sec:validation} validates the code against analytic limits, spectral shapes, Green's function approximations, and tabulated Green's functions provided by \cosmotherm{}.
Section~\ref{sec:dark_photon} applies the code to dark photon oscillations and derives FIRAS constraints on kinetic mixing.
Section~\ref{sec:limitations} discusses current limitations and future directions, and Section~\ref{sec:ai} documents the human--AI development process in detail.

While AI tools were used extensively in development and drafting, the human authors are solely responsible for the correctness of the code, the physics, and all claims made in this paper.

\section{Physics of spectral distortions}
\label{sec:physics}
The early universe contains a hot, dense plasma of photons, electrons, and ions that interact through several scattering processes.
Compton scattering ($e^- \gamma \to e^- \gamma$) exchanges energy between photons and electrons but conserves photon number, while double Compton (DC) and Bremsstrahlung (BR) emission create and destroy photons.
When all these processes are faster than the Hubble expansion ($z \gtrsim 2\times 10^6$), any injected energy is rapidly thermalized into a slightly hotter blackbody.
When they are slower, the energy leaves a lasting imprint: a spectral distortion whose shape encodes when and how the energy was injected (Sec.~\ref{sec:thermalization_regimes}).
Energy injection heats the electrons above the photon bath ($\Te > \Tz$), and this temperature difference drives energy transfer into the photon field through Compton scattering.

In this section we describe the physics that \spectroxide{} solves to predict these spectral distortions. 
We provide an overview of the photon Boltzmann equation with Compton, double Compton, and Bremsstrahlung scattering (Sec.~\ref{sec:boltzmann}), closely following the formulation of Ref.~\cite{Chluba2012cosmotherm}.
Throughout, we work in natural units ($k_B = c = \hbar = 1$) and follow the notation of Ref.~\cite{Chluba2012cosmotherm}: $x = 2\pi\nu / T_\gamma$ where $T_\gamma = T_0(1+z)$ is a reference temperature defined by the present-day CMB monopole $T_0 = \qty{2.7255}{\kelvin}$~\cite{Fixsen2009}, $\te = \Te / m_\mathrm{e}$, and $\theta_\gamma = T_\gamma / m_\mathrm{e}$. It is also important to clarify that in the presence of spectral distortions, $T_\gamma$ is not a true thermodynamic temperature but a fixed reference scale against which the distortion is measured.

\subsection{Boltzmann equation and scattering processes}
\label{sec:boltzmann}

In this work, we focus on small deviations from the Planck blackbody spectrum, which allows us to decompose the photon occupation number as $n(x, \tau) = \npl(x) + \dn(x, \tau)$, where $\npl(x) = 1/(e^x - 1)$ is the Planck distribution at temperature $\Tz$ and $\dn \ll \npl$ is the spectral distortion, and we have expressed the time variable in terms of the Thomson optical depth $d\tau = n_\mathrm{e} \sigma_\mathrm{T} \, dt$, where $\sigma_\mathrm{T}$ is the Thomson cross section.
Throughout, we use the spectral moments
\begin{equation}
    G_k \equiv \int x^k\,n\,dx\,,\qquad
    I_k \equiv \int x^k\,n(1{+}n)\,dx\,,    \label{eq:moments}
\end{equation}
evaluated on the full occupation $n = \npl + \dn$.
A superscript $(0)$ denotes the Planck value, so that $\Gpl{2} = 2\zeta(3)$, $\Gpl{3} = \pi^4/15$, and $\Ipl{4} = 4\pi^4/15$.
The evolution of $n$ is governed by the Boltzmann equation, written in terms of Thomson optical depth  as
\begin{equation}
    \frac{\partial n}{\partial \tau} = \left.\frac{\partial n}{\partial \tau}\right|_\mathrm{K} + \left.\frac{\partial n}{\partial \tau}\right|_\dc + \left.\frac{\partial n}{\partial \tau}\right|_\br + S(x, \tau)\,,    \label{eq:boltzmann}
\end{equation}
where the terms on the right represent Compton scattering, double Compton emission, Bremsstrahlung, and any external photon source $S(x,\tau)$.
In this paper, we consider two different injection mechanisms: 1) heat injection, where energy is added to the baryons, leading to $S = 0$ and the energy transfer to the CMB entering indirectly through an increase in $T_\mathrm{e}$, and 2) photon injection, where photons are added directly to the spectrum through a nonzero $S(x,\tau)$.
We summarize the terms in Eq.~\eqref{eq:boltzmann} below; the full equations, reformulated in terms of $\dn$, are given in Appendix~\ref{app:scattering}.

Compton scattering redistributes photons in frequency and its precise effect depends on the difference between $\Te$ and $\Tz$.
When $\Te > \Tz$, electrons up-scatter photons, transferring energy into the photon field.
Because Compton scattering conserves photon number, it can establish a Bose--Einstein distribution but cannot relax the chemical potential $\mu$ to zero, which instead requires DC and BR.

DC ($\gamma e^- \!\to\! \gamma\gamma e^-$) and BR ($e^- X^{+Z} \!\to\! e^- X^{+Z} \gamma$) create and destroy photons, providing the $\mu$-relaxation that Compton scattering alone cannot achieve~\cite{Lightman1981,Chluba2007dc,Rybicki1979}.
Both share a common rate structure (Eq.~\eqref{eq:dcbr}) with emission coefficients $K$ that diverge as $1/x^3$ at low frequencies, causing sufficiently soft photons to be absorbed on timescales far shorter than the Hubble time.
DC dominates the photon-number-changing rate at $z \gtrsim 10^6$ due to its steep temperature scaling ($K_\dc \propto \tz^2$), while BR dominates at $z \lesssim 10^5$; the two contribute comparably in the intermediate decade.
The stiffness of these rates, combined with the Compton near-cancellation, makes the system difficult to integrate (Appendix~\ref{app:numerics}).

The three scattering processes are coupled through the electron temperature $\Te$.
Physically, $\Te$ sits at a quasi-stationary balance between heating and cooling effects. Processes like DC/BR photon absorption and external energy injection heat the electrons. Simultaneously, Compton scattering transfers energy between the electron and photon fluids, pushing the two fluids toward thermal equilibrium. 
Because the Compton scattering energy transfer rate is large before recombination, $\Te$ adjusts nearly instantaneously to the current $\dn$ and heating rate. 

With the scattering rates and $\Te$ determined by the photon distribution, the remaining computational task is to integrate the Boltzmann equation forward in redshift.
\spectroxide{} evolves $\dn(x, z)$ on a non-uniform frequency grid from an initial redshift $z_\mathrm{start}$ (chosen to capture the full injection history) down to the present day.
At each redshift step, the solver: (i) updates the free electron fraction $X_\mathrm{e}(z)$ from a Peebles three-level atom model~\cite{Peebles1968,Zeldovich:1969qmq}, evaluated at the current $\Te$ and with the corrections of RECFAST~\cite{Seager1999,Rubino-Martin2010} that bring it to within 0.4\% of the multi-level code HyRec~\cite{AliHaimoud2011,Lee:2020obi}; (ii) determines $\Te$ self-consistently from the current $\dn$ and heating rate; and (iii) advances $\dn$ by one coupled Compton + DC/BR time step using the scheme described in Appendix~\ref{app:numerics}.
The output is $\dn(x)$ at the present day, from which we extract the observable distortion parameters $\mu$, $y$, and $\Delta T/T$, which characterize the shape of a spectral distortion, as we explain in more detail in Sec.~\ref{sec:thermalization_regimes}. 

The shape of the resulting distortion depends on the injection redshift, because the efficiency of the scattering processes changes as the universe expands and cools; we describe the three relevant regimes and their characteristic spectral shapes in Sec.~\ref{sec:thermalization_regimes}.

\section{Heat injection}
\label{sec:heat_injection}
For heat injection, no photons are added to the spectrum directly ($S = 0$ in Eq.~\eqref{eq:boltzmann}).
Instead, the injected energy heats the electrons above the photon bath, raising $\re = \Te/\Tz$ above unity.
Compton scattering then transfers this excess energy into the photon field, creating the spectral distortion.

The same mechanism operates in reverse even when nothing is injected.
Hubble expansion cools the electrons as $\Te \propto (1+z)^{2}$ while the CMB cools as  $\Tz \propto (1+z)$, so $\re$ falls slightly below unity even before thermal decoupling of the electrons from the CMB, and Compton scattering drains energy out of the photon field.
\spectroxide{} carries this adiabatic-cooling term in the $\Te$ evolution equation (Appendix~\ref{app:te_eq}) and therefore computes the guaranteed $\Lambda$CDM cooling distortion automatically~\cite{Chluba:2016bvg}.
As a result, a run with no heat or photon injection returns a nonzero distortion and an injected signal with $\drho \sim 10^{-8}$ must be measured differentially, by subtracting an otherwise identical no-injection run (we do exactly this for the visibility fit of Sec.~\ref{sec:heat_validation}).

We first describe how the thermalization regime determines the spectral shape (Sec.~\ref{sec:thermalization_regimes}), then specify the injection scenarios built into \spectroxide{} and provide an example of how to implement custom injection scenarios.

\subsection{Thermalization regimes and spectral shapes}
\label{sec:thermalization_regimes}
The spectral shape of a distortion depends on \emph{when} the energy is injected, because the efficiency of the scattering processes described in Sec.~\ref{sec:boltzmann} changes with redshift.
We describe the relevant regimes below, from earliest to latest~\cite{Sunyaev1970mu,Hu1993,Chluba2012dissipation,Chluba2013greens}.

\paragraph{Thermalization era ($z \gtrsim 2 \times 10^6$).}
At sufficiently early times, DC and BR adjust the photon number on timescales shorter than both the Hubble time and the Compton time.
Injected energy is fully absorbed: energy and photon number both relax to equilibrium, producing a blackbody at a slightly higher temperature.
To leading order, the difference in occupation number between two blackbodies at temperatures $T$ and $T + \Delta T$ is $\dn = (\Delta T/T)\,\Gbb(x)$, where
\begin{equation}
    \Gbb(x) = \frac{x\,e^x}{(e^x - 1)^2}    \label{eq:gbb}
\end{equation}
is the temperature-shift spectral shape and $\Delta T / T$ is its amplitude.
Since FIRAS measures the CMB temperature as a free parameter, this component is unobservable~\cite{Fixsen1996,Chluba2012cosmotherm,CJ2014} and injections in the thermalization era therefore leave no detectable imprint on the spectrum.
It is important to note that this picture applies to small distortions where $\drho \ll 1$; for larger energy releases, nonlinear effects modify the thermalization process~\cite{Chluba2012cosmotherm,Khatri2012}, which we do not model in this work.

More generally, any spectral distortion $\dn(x)$ contains an unobservable temperature-shift component proportional to $\Gbb(x)$.
Following the convention of Refs.~\cite{Chluba2012cosmotherm,Chluba2013greens}, we subtract this component from the raw PDE output $\dn$ by computing
\begin{equation}
    \alpha_T = \frac{\int x^2\,\dn\,dx}{\int x^2\,\Gbb\,dx}\,,\qquad \dn \to \dn - \alpha_T\,\Gbb\,,
\end{equation}
which enforces $\int x^2\,\dn\,dx = 0$.
This number-conserving (NC) projection removes the unobservable temperature-shift component from all PDE spectra.

\paragraph{$\mu$ era ($3 \times 10^5$$\, \lesssim z \lesssim 2 \times 10^6$).}
As the universe expands and cools, DC and BR become too slow to maintain full thermal equilibrium, but Compton scattering remains efficient and drives the photon distribution toward a Bose--Einstein distribution instead, with a potentially nonzero chemical potential.
Because Compton scattering conserves photon number but redistributes energy, the equilibrium temperature generally differs from $\Tz$, absorbing part of the injected energy as an unobservable temperature shift.
Expanding to first order in $\Delta T/\Tz$ and $\mu$ gives
\begin{equation}
    n_\mathrm{BE} - \npl \approx \left(\frac{\Delta T}{\Tz} - \frac{\mu}{\bmu}\right)\Gbb(x) \;+\; \mu\,M(x)\,,    \label{eq:be_expansion}
\end{equation}
where the first term is a temperature shift (unobservable, as discussed above) and
\begin{equation}
    M(x) = \left(\frac{x}{\bmu} - 1\right) \frac{e^x}{(e^x - 1)^2}    \label{eq:mshape}
\end{equation}
is the observable $\mu$-distortion spectral shape, with $\bmu = 3\zeta(3)/\zeta(2) \approx 2.192$ fixed by photon number conservation ($\int x^2 M\,dx = 0$).
After stripping the unobservable temperature shift, the residual distortion is $\dn \approx \mu\,M(x)$.
The relationship between $\mu$ and the fractional energy injection is
\begin{equation}
    \mu = \frac{3}{\kc}\,\drho \approx 1.401\,\drho\,,
    \label{eq:mu_drho}
\end{equation}
where $\kc = 3\int x^3 M(x)\,dx / \Gpl{3} \approx 2.142$~\cite{Sunyaev1970mu,Illarionov1975}.

\paragraph{Intermediate regime ($10^4$$\, \lesssim z \lesssim\,$$3 \times 10^5$).}
The distortion during this regime cannot be described by a single analytic template; only the full PDE resolves the spectral shape accurately.

\paragraph{$y$ era ($z \lesssim\,$$10^4$).}
At late times, Compton scattering is too slow to redistribute photon energies, and the distortion retains the Compton-$y$ shape:
\begin{equation}
    \Ysz(x) = \frac{x\,e^x}{(e^x - 1)^2} \left( x\,\frac{e^x + 1}{e^x - 1} - 4 \right)\,,    \label{eq:yshape}
\end{equation}
with $\dn \approx y\,\Ysz(x)$.
The $y$-parameter is related to the energy injection by $y = \drho / 4$.

The procedure for extracting $\mu$, $y$, and $\Delta T/T$ from a numerical $\dn(x)$ is described in Appendix~\ref{app:decomposition}.

\subsection{Injection scenarios}
\label{sec:heat_scenarios}

Any user-defined heat injection scenario can be specified to \spectroxide{}, but single bursts, decaying particles, and annihilating dark matter are already built into the code.
Each scenario specifies the fractional heating rate $d(\drho)/dz$ that enters the electron temperature evolution through $\delta{\re}_\mathrm{inj}$ (Eqs.~\eqref{eq:te_ode}--\eqref{eq:te_params}).
Photon injection, which enters through the source term $S(x,\tau)$ rather than through the heating rate, is treated separately in Sec.~\ref{sec:photon_injection}. 

\paragraph{Single burst.}
A quasi-instantaneous injection at redshift $z_h$ with total energy $\drho$, implemented as a temporal Gaussian with width $\sigma_z$ (default $\sigma_z = \max(0.04\,z_h, 100)$):
\begin{equation}
    \frac{d(\drho)}{dz} = \frac{\drho}{\sqrt{2\pi}\,\sigma_z} \exp\!\left[-\frac{(z - z_h)^2}{2\sigma_z^2}\right]\,.\end{equation}
The width of the injection is chosen to be narrow enough that the injection is approximately instantaneous, but not so narrow that the solver cannot resolve it.
We have checked that the default $\sigma_z$ accomplishes both these goals, but the user can modify this default if they wish.

\paragraph{Decaying particle.}
A massive particle $X$ with mass $m_X$ and lifetime $\tau_X = \Gamma_X^{-1}$ (not to be confused with the Thomson optical depth $\tau$) decays and deposits a fraction $\epsilon_X$ of its rest mass energy as heat.
The volumetric energy injection rate is simply the energy per decay times the decay rate per unit volume:
\begin{equation}
    \frac{dE}{dt\,dV}\bigg|_\mathrm{dec} = \epsilon_X \, m_X \, n_X(z) \, \Gamma_X \, e^{-\Gamma_X t}\,,    \label{eq:dec_fund}
\end{equation}
where $n_X(z) = n_{X,0}(1+z)^3$ is the particle number density and the exponential accounts for the depletion of the parent population over time.

Following the approach of Refs.~\cite{Chluba2012cosmotherm,Chluba2013DM}, we can rewrite this rate  by defining 
\begin{equation}
    f_X \equiv \epsilon_X\,m_X\,n_{X,0}/n_{\mathrm{H},0}\,,
\end{equation}
which has units of \unit{\eV} and represents the total energy released per hydrogen atom if $X$ were to decay fully. This definition allows us to more clearly separate the model-dependent particle-physics dependence from the cosmology dependence, although we note that the remaining particle physics parameter $\Gamma_X$ is not absorbed in $f_X$. Here, $n_{\mathrm{H,0}}$ is the present-day hydrogen number density given by $n_\mathrm{H,0} = (1 - Y_p)\,n_{b,0}$, where $Y_p$ is the primordial helium-4 mass fraction and $n_{b,0} = \rho_{b,0}/m_p$ is the total baryon number density (counting each helium-4 nucleus as four baryons).
Then, we can rewrite Eq.~\eqref{eq:dec_fund} as 
\begin{equation}
    \frac{dE}{dt\,dV}\bigg|_\mathrm{dec} = f_X \, \Gamma_X \, n_\mathrm{H}(z) \, e^{-\Gamma_X t}\,.    \label{eq:decay_rate}
\end{equation}

To get the fractional heating rate that enters \spectroxide{}, we divide by the photon energy density $\rho_\gamma$ and convert from cosmic time to redshift:
\begin{equation}
    \frac{d(\drho)}{dz} = \frac{f_X \Gamma_X n_\mathrm{H}(z)}{H(z)\rho_\gamma(z)(1+z)}e^{-\Gamma_Xt} \, .\end{equation}

\paragraph{Dark matter annihilation.}
For $s$-wave annihilation with thermally averaged cross section $\langle\sigma v\rangle$, the volumetric energy injection rate is
\begin{equation}
    \frac{dE}{dt\,dV}\bigg|_\mathrm{ann} = f_\mathrm{eff}\,\langle\sigma v\rangle\,m_\chi\,n_\chi^2(z)\,,    \label{eq:ann_fundamental}
\end{equation}
where $f_\mathrm{eff}$ is the fraction of annihilation energy deposited as heat, $m_\chi$ is the DM particle mass, and $n_\chi(z) = (\rho_\mathrm{cdm,0}/m_\chi)(1+z)^3$ is the DM number density.
This expression is correct for self-conjugate DM, but the extra factor of $1/2$ that enters for Dirac fermions can simply be absorbed into $f_\mathrm{eff}$. 
The rate scales as $n_\chi^2 \propto (1+z)^6$.

As with decays, we absorb the particle physics into a single parameter~\cite{Chluba2012cosmotherm,Chluba2013DM}:
\begin{equation}
    f_\mathrm{ann} \equiv f_\mathrm{eff}\,\langle\sigma v\rangle\,m_\chi\,\frac{n_{\chi,0}^2}{n_{\mathrm{H},0}}\,,     \label{eq:fann_def}
\end{equation}
which has units of \unit{\eV\per\second}, so that the rate becomes $dE/(dt\,dV) = f_\mathrm{ann}\,n_\mathrm{H}(z)\,(1+z)^3$.
The fractional heating rate is
\begin{equation}
    \frac{d(\drho)}{dz}\bigg|_\mathrm{ann} = \frac{f_\mathrm{ann}\,n_\mathrm{H}(z)\,(1+z)^3}{H(z)\,\rho_\gamma(z)\,(1+z)} \propto (1+z)^{-1}\,,    \label{eq:drho_dz_ann}
\end{equation}
where the scaling follows from $n_\mathrm{H}(1+z)^3 \propto (1+z)^6$, $\rho_\gamma \propto (1+z)^4$, and $H \propto (1+z)^2$ during radiation domination.

For $p$-wave annihilation, the cross section acquires a velocity dependence $\langle\sigma v\rangle \propto T \propto (1+z)$~\cite{McDonald2001,Chluba2013DM} for DM thermally coupled to the CMB. This adds one power of $(1+z)$ to the heating rate relative to $s$-wave~\cite{Chluba2013DM}:
\begin{equation}
    \frac{d(\drho)}{dz}\bigg|_\mathrm{p\text{-}wave} = \frac{f_\mathrm{ann}\,n_\mathrm{H}(z)\,(1+z)^4}{H(z)\,\rho_\gamma(z)\,(1+z)} \propto \mathrm{const}\,.    \label{eq:pwave_rate}
\end{equation}
Here, $f_\mathrm{ann}$ retains the same definition as in Eq.~\eqref{eq:fann_def} but now includes the present-day value (assuming that $T \propto (1 + z)$ at all times) of the velocity-dependent cross section.
The flatter redshift scaling compared to $s$-wave shifts the energy injection toward higher redshifts, producing a more $\mu$-dominated spectrum.
$p$-wave annihilation for nonrelativistic DM that is not thermally coupled to the CMB can be implemented as a custom scenario, but is generically less constrained by spectral distortions due to the less favorable redshift scaling, $\langle \sigma v\rangle \propto (1+z)^2$. 

\paragraph{Custom scenarios.}
\label{sec:custom_scenarios}
For injection histories not covered by the built-in scenarios, \spectroxide{} accepts an arbitrary user-defined heating rate $d(\drho)/dz$ as a Python function.
This is the primary interface for connecting new-physics models to the spectral distortion machinery: any model that predicts an energy injection rate as a function of redshift can be passed directly to the solver without modifying the code.

As an example, we provide an implementation of a sinusoidal heating/cooling profile that spans the $\mu$--$y$ transition (one of the pathological stress tests from Sec.~\ref{sec:validation}):
\begin{lstlisting}
import numpy as np
import spectroxide

def dq_dz(z):
    """Sinusoidal heating/cooling over mu-y transition."""
    k = 2 * np.pi / 5e4
    return 5e-10 * np.sin(k * (1 + z))

# Full PDE solution
result = spectroxide.solve(method="pde", dq_dz=dq_dz,
                           z_start=5e5, z_end=1e3)
\end{lstlisting}
The \code{dq\_dz} function must return $d(\drho)/dz$ (positive for heating, negative for cooling) and should integrate to the desired total $\drho$ over the injection epoch.
There are no restrictions on the functional form: the solver handles oscillatory profiles, sharp peaks, broad power laws, and profiles with regions of net cooling, as demonstrated by the stress tests in Sec.~\ref{sec:validation}.
The returned \code{result} object contains the frequency grid $x$, the spectral distortion $\dn(x)$, and the distortion parameters $\mu$ and $y$ extracted via the least-squares decomposition described in Appendix~\ref{app:decomposition}.

\section{Photon injection}
\label{sec:photon_injection}

Unlike heat injection, where energy enters indirectly through $\Te$, photon injection adds photons directly to the spectrum through the source term $S(x, \tau)$ in Eq.~\eqref{eq:boltzmann}.
The fate of these injected photons depends on their frequency relative to a critical frequency $x_c(z)$ where the DC/BR optical depth reaches unity.
Since DC/BR rates scale as $1/x^3$, absorption is extremely efficient at low frequencies: $x_c \sim 0.006$ at $z \sim 10^6$.
At frequencies well below $x_c$, the coupled DC/BR solver absorbs the photons within a single time step; the absorbed energy raises $\Te$ through the DC/BR back-reaction integral $H_{\dc/\br}$ (Eq.~\eqref{eq:hdcbr}) and is returned to the photon field through the Compton operator.
At higher frequencies ($x \gg x_c$), DC/BR absorption is negligible and the injected photons can survive as a localized spectral feature.
Compton scattering redistributes this feature: in the $\mu$-era ($z \gtrsim\,$$3 \times 10^5$), Comptonization is efficient and rapidly thermalizes the excess into a $\mu$-distortion, while in the $y$-era ($z \lesssim\,$$10^4$), Comptonization is slow and the feature broadens but remains as a distinct bump.
This interplay between survival and absorption produces a rich phenomenology that depends on both the injection frequency and redshift.

We describe the implemented injection scenarios below (Sec.~\ref{sec:photon_scenarios}), including monochromatic bursts, decaying particle photon emission, and user-defined sources.
Validation of the photon injection solver is presented in Sec.~\ref{sec:photon_validation}.

\subsection{Injection scenarios}
\label{sec:photon_scenarios}

\paragraph{Monochromatic injection.}
The simplest photon injection scenario is a monochromatic burst at dimensionless frequency $x_\mathrm{b}$ and redshift $z_h$, injecting a fractional number density $\Delta N / N$ of photons, where $N$ is the total number density of CMB photons.
The corresponding source term is~\cite{Chluba2015photon}
\begin{equation}
    S(x, \tau) = \frac{\Delta N}{N}\,\frac{\Gpl{2}}{x^2}\,\mathcal{G}(x; x_\mathrm{b}, \sigma_x)\,\mathcal{G}(z; z_h, \sigma_z)\,\left|\frac{dz}{d\tau}\right|\,,
    \label{eq:photon_source}
\end{equation}
where $\mathcal{G}(u; u_0, \sigma) = (2\pi\sigma^2)^{-1/2}\exp[-(u-u_0)^2/(2\sigma^2)]$ denotes a normalized Gaussian and $|dz/d\tau| = H(z)(1+z)/(n_\mathrm{e}\,\sigma_\mathrm{T})$ converts the temporal Gaussian from per-redshift to per-Thomson-optical-depth units.
The structure of this expression is straightforward: the two Gaussians localize the injection in frequency and time, while the prefactor $\Gpl{2}/x^2$ sets the overall normalization.
Because the photon number density scales as $\int x^2\,n\,dx$, the occupation number contributed by $\Delta N/N$ photons at a single frequency goes as $1/x^2$. The factor $\Gpl{2}$ then ensures $\int x^2\,S\,d\tau\,dx = (\Delta N/N)\,\Gpl{2}$, so that the total fractional photon number injected is exactly $\Delta N/N$.
The frequency width $\sigma_x = 0.05\,x_\mathrm{inj}$ is chosen narrow enough to approximate a monochromatic line while remaining resolvable on the frequency grid, and the temporal width $\sigma_z$ defaults to $\max(0.04\,z_h, 100)$, matching the single-burst energy injection convention.

\paragraph{Decaying particle photon injection.}
For a particle $X$ decaying into photon pairs ($X \to \gamma\gamma$), each photon carries energy $m_X/2$, corresponding to a redshift-dependent normalized injection frequency $x_\mathrm{inj}(z) = m_X / (2\Tz)$.
The injection rate follows the exponential decay profile of Eq.~\eqref{eq:decay_rate}, but the energy is deposited as photons at $x_\mathrm{inj}(z)$ rather than as heat, following Ref.~\cite{Bolliet2021}.
Only spontaneous emission is included; stimulated emission is neglected.
Because $X \to \gamma\gamma$ deposits both final-state photons into the same mode at $x_\mathrm{inj}$, the full rate carries a Bose enhancement factor $(1 + \npl(x_\mathrm{inj}))^2$---one $(1+\npl)$ per emitted photon, squared because both photons occupy the same state (for single-photon emission channels the factor would instead be $(1+\npl)$).
The omission is accurate for $x_\mathrm{inj} \gg 1$ where $\npl$ is exponentially small; for $x_\mathrm{inj} \lesssim 1$ the $(1+\npl)^2$ correction becomes significant ($\npl \approx 1/x_\mathrm{inj}$, so the missing factor grows as $1/x_\mathrm{inj}^2$) and should be included via the custom photon source interface described below.
More generally, electromagnetic cascades (pair production, photon splitting), interactions with nuclei, and secondary photon production are not modeled; these effects become important for high-energy injection ($\omega \gtrsim m_\mathrm{e}$) and are left to future work.

\paragraph{Dark photon oscillations.}
\spectroxide{} also includes a built-in dark photon oscillation scenario, which acts as a frequency-dependent photon \emph{depletion} source: resonant $\gamma \to A'$ conversion removes photons from the CMB spectrum at a redshift set by the dark photon mass~\cite{Mirizzi2009,Caputo2020,Chluba2024darkphoton}.
This uses the same photon source infrastructure but with a negative source term.
The physics and FIRAS constraints are discussed in detail in Sec.~\ref{sec:dark_photon}.

\paragraph{Custom photon sources.}
For photon injection profiles not covered by the built-in scenarios, \spectroxide{} accepts a user-defined photon source function 
\begin{equation}
    f(x, z) \equiv \frac{d(\dn(x, z))}{dz},
    \label{eq:custom_photon_source}
\end{equation}
the occupation-number injection rate per unit redshift at each dimensionless frequency $x$---the frequency-dependent generalization of the scalar heating rate $d(\drho)/dz$ used for custom heat injection.
The solver internally converts to the Boltzmann source term $S(x, \tau)$ in Eq.~\eqref{eq:boltzmann}.

As an example, the following extends the built-in decaying particle source (Eq.~\eqref{eq:decay_rate}) with Bose-enhanced stimulated emission.
The injection frequency $x_\mathrm{inj}(z) = m_X/(2\Tz)$ redshifts as $1/(1+z)$ since the photon energy is fixed at $m_X/2$ while $\Tz \propto (1+z)$.
The Gaussian line profile and $1/x^2$ occupation-number weighting follow from Eq.~\eqref{eq:photon_source}; the only new ingredient is the Bose enhancement factor $(1 + \npl(x))^2$ appropriate for the two-photon $X \to \gamma\gamma$ channel (both photons into the same mode at $x_\mathrm{inj}$), which amplifies the emission rate into occupied modes and can be significant for low-frequency injection ($x_\mathrm{inj} \lesssim 1$, where $\npl \gg 1$).
The overall amplitude encodes the particle abundance and decay rate; we use a representative value here for illustration:
\begin{lstlisting}
import spectroxide
import numpy as np

m_X = 10.0     # particle mass [eV]
T0  = 2.35e-4  # CMB temperature today [eV]

def stimulated_decay(x, z):
    """X -> gamma gamma with Bose-enhanced emission."""
    x_inj = m_X / (2 * T0 * (1 + z))
    sig = 0.05 * x_inj
    line = np.exp(-0.5*((x - x_inj)/sig)**2) \
           / (sig * np.sqrt(2*np.pi))
    bose = (1 + 1 / (np.exp(x) - 1))**2
    return 6e-11 * line / x**2 * bose

result = spectroxide.solve(method="pde",
    photon_source=stimulated_decay,
    z_start=2e5, z_end=2e4)
\end{lstlisting}
As with heat injection, the returned \code{result} object contains the frequency grid $x$, the spectral distortion $\dn(x)$, and the distortion parameters $\mu$, $y$, and $\drho$.

\section{Validation}
\label{sec:validation}

Having described the physics of heat injection (Sec.~\ref{sec:heat_injection}) and photon injection (Sec.~\ref{sec:photon_injection}), we now validate \spectroxide{} at progressively more demanding levels.
We first introduce the Green's function formalism as an independent semi-analytic cross-check (Sec.~\ref{sec:greens}), then validate heat injection against analytic limits, Green's functions tabulated by \cosmotherm{} v1.0.3~\cite{Chluba2012cosmotherm}, and pathological stress tests (Sec.~\ref{sec:heat_validation}), followed by photon injection validation (Sec.~\ref{sec:photon_validation}).
Dark photon oscillation constraints, which serve as an end-to-end application, are presented separately in Sec.~\ref{sec:dark_photon}.
The code repository includes Jupyter notebooks with extended validation beyond what is presented here, which we invite users to explore.
Full details of the numerical methods (time integration, frequency grid, adaptive stepping, energy conservation, and convergence tests) are given in Appendix~\ref{app:numerics}. Numerical convergence and energy conservation tests are presented in Appendices~\ref{app:energy} and~\ref{app:convergence}.
This section validates the physics; we provide additional details of code \emph{verification} (the process by which we check that the discretized equations are solved correctly) in Appendices~\ref{app:mms} and~\ref{app:moments}.

\subsection{Green's function formalism}
\label{sec:greens}
A powerful independent check on the PDE solver comes from the Green's function formalism, which provides semi-analytic predictions for both energy injection~\cite{Chluba2013greens} and photon injection~\cite{Chluba2015photon,Arsenadze2025}.
We describe the formalism here and use it to validate the PDE solver for heat injection (Sec.~\ref{sec:heat_validation}) and photon injection (Sec.~\ref{sec:photon_validation}).

\subsubsection{Heat injection Green's function}
\label{sec:heat_gf}

For small distortions, the Boltzmann equation Eq.~\eqref{eq:boltzmann} is approximately linear in $\dn$, so the spectral response to any continuous energy injection history $d(\drho)/dz$ can be written as a superposition of responses to instantaneous bursts:
\begin{equation}
    \dn(x) = \int_{z_\mathrm{min}}^{z_\mathrm{max}} \Gth(x, z)\,\frac{d(\drho)}{dz}\,dz\,.
    \label{eq:convolution}\end{equation}
Here, $\Gth(x, z)$ is the thermalization Green's function introduced by Ref.~\cite{Chluba2013greens} and is the spectral distortion at frequency $x$ produced by a unit delta function energy injection at redshift $z$.

Refs.~\cite{Chluba2013greens,Chluba2015photon} showed that the Green's function can be approximated analytically by decomposing it into the three spectral shapes from Sec.~\ref{sec:thermalization_regimes}:
\begin{align}
    \Gth(x, z_h) &= \frac{3}{\kc}\,\Jmu(z_h)\,\Jbb(z_h)\,M(x) \nonumber \\
                  &\quad + \frac{\Jy(z_h)}{4}\,\Ysz(x) \nonumber \\
                  &\quad + \frac{1 - \Jbb(z_h)}{4}\,\Gbb(x)\,,     \label{eq:gth}
\end{align}
where $z_h$ is the injection redshift and $\Jbb(z)$, $\Jmu(z)$, and $\Jy(z)$ are visibility functions that govern the branching between thermalization, $\mu$-type, and $y$-type distortions.
Fitting formulas for these visibility functions---and their parameter values, both from the literature and from fits to \spectroxide{} data---are presented in Sec.~\ref{sec:heat_validation} below (Eqs.~\eqref{eq:jbb}--\eqref{eq:jy}, Table~\ref{tab:visibility_params}, Fig.~\ref{fig:visibility}).

Eq.~\eqref{eq:gth} is an inherently limited approximation at intermediate redshifts ($10^4$$\, \lesssim z_h \lesssim\,$$3 \times 10^5$), where the true spectral shape is non-trivial and cannot be captured by a linear combination of $M$ and $\Ysz$, but it provides a useful approximation~\cite{Chluba2013greens}.

Besides this numerical approximation, a table of Green's functions $G(x,z_i)$ for individual $z_i$ can also be computed numerically by injecting a single burst of energy at $z_i$ and computing the resulting distortion $\Delta n(x)$ using \spectroxide{} or another PDE solver. Then, using Eq.~\eqref{eq:convolution}, the table of Green's functions can be convolved with an arbitrary energy injection history to acquire the resulting spectral distortion.

We include the functionality to produce such a table of Green's functions and convolve them with custom injection histories in \spectroxide. Given the speed of the main PDE solver, this is generally unnecessary, but it provides an additional validation method for the PDE solver by comparing our Green's functions with analytic expressions (e.g.\ in Eq.~\eqref{eq:gth} for heat injection and as given in the next section for photon injection), or with the publicly available \cosmotherm\ Green's functions for heat injection~\cite{Chluba2012cosmotherm,Chluba2013greens}.

\subsubsection{Photon injection Green's function}
\label{sec:photon_gf}

Photon injection differs from heat injection because the injected photons carry both energy and number, and the resulting distortion depends on the injection frequency.
Ref.~\cite{Chluba2015photon} developed a Green's function approximation for this case, analogous to Eq.~\eqref{eq:gth}.
Ref.~\cite{Chluba2015photon} defines the photon Green's function in intensity space; we instead work with the occupation-number distortion, related by $\Delta I_\nu = 4\pi\nu^3\,\dn(x)$, and absorb this conversion into the Green's function below. The spectral distortion from an arbitrary photon source is
\begin{equation}
    \dn(x) = \int dz \int d\nu'\;G_\gamma(x;\,\nu',\,z)\;\frac{d^2({\Delta N}/{N})}{d\nu'\,dz}(\nu',\,z)\,,
    \label{eq:photon_gf}
\end{equation}
where $d^2(\Delta N/N)/(d\nu'\,dz)$ is the fractional photon number injected per unit frequency per unit redshift and $G_\gamma(x;\,\nu',\,z)$ is the photon injection Green's function (in occupation-number form).

The photon injection Green's function $G_\gamma$ encodes how injected photons are reprocessed by Compton scattering, DC emission, and BR---the same processes that govern heat injection, but now acting on a localized spectral injection at a particular frequency and a particular redshift.
A key ingredient is the competition between injection and DC/BR absorption.
At low frequencies the DC/BR absorption rate $K/x^3$ is large, so injected photons are rapidly absorbed.
When these processes are efficient, this then defines a critical frequency $x_c(z)$ below which the DC/BR optical depth exceeds unity.
Following Ref.~\cite{Chluba2015photon}, the individual DC and BR critical frequencies scale as
\begin{equation}
    x_{c,\dc}(z) \approx 8.60 \times 10^{-3}\!\left(\frac{1+z}{2\times 10^6}\right)^{\!1/2}\!, \quad
    x_{c,\br}(z) \approx 1.23 \times 10^{-3}\!\left(\frac{1+z}{2\times 10^6}\right)^{\!-0.672}\!,
    \label{eq:xc}
\end{equation}
with the combined critical frequency $x_c^2 = x_{c,\dc}^2 + x_{c,\br}^2$ (quadrature addition, following Ref.~\cite{Chluba2015photon} Eq.~25).

The probability $P_s(x_\mathrm{inj}, z)$ that an injected photon survives DC/BR absorption has two regimes~\cite{Chluba2015photon}:
\begin{equation}
    P_s(x_\mathrm{inj}, z) =
    \begin{cases}
        \exp\!\bigl[-x_c(z)/x_\mathrm{inj}\bigr] & \mu\text{-era}\,,\\[4pt]
        \exp\!\bigl[-\tau_{\mathrm{ff}}(x_\mathrm{inj}, z)\bigr] & y\text{-era}\,,
    \end{cases}
    \label{eq:ps}
\end{equation}
where $\tau_{\mathrm{ff}}$ is the cumulative DC/BR absorption optical depth,
\begin{equation}
    \tau_{\mathrm{ff}}(x_\mathrm{inj}, z) = \int_0^z \frac{K(x_\mathrm{inj}, z')}{x_\mathrm{inj}^3}\bigl(e^{x_\mathrm{inj}\phi(z')} - 1\bigr)\,\frac{n_e\,\sigma_T}{(1+z')\,H(z')}\,dz'\,,
    \label{eq:tau_ff}
\end{equation}
and $K = K_{\dc} + K_{\br}$ is the combined emission coefficient.
In the $y$-era Compton coupling is efficient enough to enforce $\Te \approx \Tz$ to the sub-percent level, so in practice we set $\phi(z') \equiv \Tz(z')/\Te(z') \to 1$ when evaluating $\tau_{\mathrm{ff}}$.
In the $\mu$-era, efficient Compton scattering rapidly redistributes photons away from the absorption region, so the effective optical depth is set by the instantaneous critical frequency $x_c$ (Eq.~\eqref{eq:xc}) rather than the cumulative integral.
However, in the $y$-era, Compton redistribution is negligible and the raw absorption integral applies; using the $\mu$-era formula here would underestimate the survival probability.
For soft photons ($x_\mathrm{inj} \ll x_c$), $P_s \to 0$: the photons are absorbed and their energy thermalizes into $\mu/y$ distortions exactly as in the heat injection case.
For hard photons ($x_\mathrm{inj} \gg x_c$), $P_s \to 1$: the photons survive and appear as a spectral feature near their injection frequency, broadened by Compton scattering.

In the $\mu$-era, Compton scattering is efficient enough to redistribute the injected photons into a Bose--Einstein spectrum, and the photon Green's function takes the form~\cite{Chluba2015photon}
\begin{equation}
    G_\mu(x;\,x_\mathrm{inj},\,z) = \alpha_\rho\,x_\mathrm{inj}\,
    \frac{3}{\kc}\,\Jbb(z)\Bigl(1 - P_s\,\frac{x_0}{x_\mathrm{inj}}\Bigr)\,M(x)
    + \frac{\lambda}{4}\,\Gbb(x)\,,    \label{eq:photon_gf_mu}
\end{equation}
where $\alpha_\rho\,x_\mathrm{inj} = (\Gpl{2}/\Gpl{3})\,x_\mathrm{inj}$ converts photon number to fractional energy, $x_0 = 4/(3\alpha_\rho) \approx 3.60$ is the balanced frequency at which a surviving photon carries exactly the right energy-per-photon ratio to produce zero net $\mu$, and $\lambda = \alpha_\rho\,x_\mathrm{inj}\bigl[1 - (1 - P_s\,x_0/x_\mathrm{inj})\,\Jbb(z)\bigr]$ enforces energy conservation by absorbing the residual into a temperature shift, depending on the thermalization efficiency.
Injecting below $x_0$ adds more number than energy (negative $\mu$), while injecting above $x_0$ does the opposite (positive $\mu$).

The $\mu$-era expression Eq.~\eqref{eq:photon_gf_mu} applies when Comptonization is efficient (deep $\mu$-era, where $y_\gamma \gg 1$~\cite{Chluba2013greens}); in the $y$-era a different Green's function is needed.
Between these two regimes the simple $\mu+y$ decomposition fails: residual ($r$-type) contributions from partial Comptonization become important~\cite{Chluba2015photon,CJ2014}, and neither $P_s$ branch in Eq.~\eqref{eq:ps} is quantitatively accurate. We therefore restrict the photon Green's function to the deep-$\mu$ ($z_h \gtrsim\,$$3\times 10^5$) and $y$-era ($z_h \lesssim\,$$10^4$) regimes; injection redshifts in the $\mu$--$y$ transition window must be evaluated with the PDE solver instead.
In the $y$-era, Compton scattering is too slow to fully redistribute the injected photons.
The Green's function, adapted from Ref.~\cite{Chluba2015photon} to the occupation-number convention used here, is
\begin{equation}
    G_y(x;\,x_\mathrm{inj},\,z) = \frac{\alpha_\rho\,x_\mathrm{inj}}{4}\!\left(1 - P_s\,\frac{e^{(\alpha + \beta)\,y_\gamma}}{1 + x_\mathrm{inj}\,y_\gamma}\right)\Ysz(x)
    + \frac{P_s\,\Gpl{2}\,x_\mathrm{inj}}{x^3}\,F(x;\,x_\mathrm{inj},\,z)\,,    \label{eq:photon_gf_y}
\end{equation}
where $y_\gamma(z)$ is the Compton $y$-parameter accumulated between injection and observation. In the second term, the $\Gpl{2} x_\mathrm{inj}/x^3$ prefactor arises when converting the intensity-space expression of Ref.~\cite{Chluba2015photon} to our occupation-number convention.
The first term is a smooth $y$-distortion whose coefficient accounts for the energy carried away by surviving photons; the second is the surviving photon bump, broadened by Compton scattering.
The Compton redistribution kernel $F(x;\,x_\mathrm{inj},\,z)$, which describes how a photon injected at $x_\mathrm{inj}$ is broadened to observed frequency $x$, is~\cite{Chluba2015photon}
\begin{equation}
    F(x;\,x_\mathrm{inj},\,z) = \frac{1}{x_\mathrm{inj}\sqrt{4\pi\beta\,y_\gamma}}\,\exp\!\left\{-\frac{\left[\ln\!\left(x(1/x_\mathrm{inj} + y_\gamma)\right) - \alpha\,y_\gamma\right]^2}{4\beta\,y_\gamma}\right\}\,, \label{eq:fcs}
\end{equation}
with normalization $\int F\,dx = e^{(\alpha+\beta)\,y_\gamma}/(1 + x_\mathrm{inj}\,y_\gamma)$~\cite{Chluba2015photon}.
The auxiliary functions that parametrize the Compton redistribution are~\cite{Chluba2015photon}
\begin{equation}
    \alpha(x_\mathrm{inj}, z) = \frac{3 - 2f(x_\mathrm{inj})}{\sqrt{1 + x_\mathrm{inj}\,y_\gamma}}\,,\quad
    \beta(x_\mathrm{inj}, z) = \frac{1}{1 + x_\mathrm{inj}\,y_\gamma\bigl[1 - f(x_\mathrm{inj})\bigr]}\,,    
    \label{eq:compton_aux}
\end{equation}
where
\begin{equation}
    f(x_\mathrm{inj}) = e^{-x_\mathrm{inj}}\!\left(1 + \frac{x_\mathrm{inj}^2}{2}\right)\!. 
\end{equation}
In the limit $y_\gamma \to 0$, $F$ reduces to a delta function at $x = x_\mathrm{inj}$ (free-streaming) and the smooth coefficient simplifies to $\alpha_\rho\,x_\mathrm{inj}\,(1 - P_s)/4$.

In the soft-photon limit ($P_s \to 0$), the bump vanishes and each of the $\mu$- and $y$-era expressions reduces to $\alpha_\rho\,x_\mathrm{inj}\,\Gth(x,z)$ in its regime of validity: $G_\mu$ (Eq.~\eqref{eq:photon_gf_mu}) recovers the $\mu$-era branch of $\Gth$ (where $\Jmu \to 1$, $\Jy \to 0$), while $G_y$ (Eq.~\eqref{eq:photon_gf_y}) recovers the $y$-era branch ($\Jmu \to 0$, $\Jy \to 1$)---the heat injection Green's function of Eq.~\eqref{eq:gth} scaled by the injected energy per photon.
\spectroxide{} includes a pure-Python implementation of Eqs.~\eqref{eq:photon_gf_mu}--\eqref{eq:photon_gf_y} for convenience.

\subsection{Heat injection validation}
\label{sec:heat_validation}
For heat injection, we validate the PDE solver against two independent benchmarks: the analytic Green's function fitting formulas for the visibility functions, and a full Green's function table from \cosmotherm{}.
These tests are complementary---the first checks whether the solver captures the correct thermalization eras (i.e., the correct $\mu/y$ branching as a function of $z_h$), while the second tests the full spectral shape at each frequency.

\begin{figure}[!t]
\centering
\includegraphics[width=\textwidth]{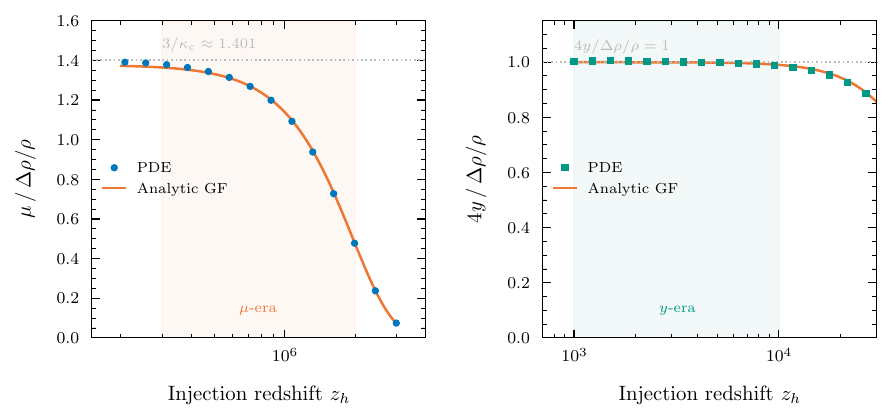}
\caption{Distortion amplitudes as a function of injection redshift.
Left: $\mu/\drho$ in the $\mu$-era and thermalization regime; the PDE (blue circles) recovers the analytic limit $3/\kc \approx 1.401$ (dotted line) to within a few percent, with DC/BR thermalization progressively suppressing $\mu$ above $z_h \sim 10^6$.
Right: $4y/\drho$ in the $y$-era and transition; the PDE (teal squares) reproduces the analytic limit of unity to within $1.2\%$ for $z_h \lesssim 10^4$.
Solid orange lines show the visibility function predictions $\Jbb\,\Jmu$ (left) and $\Jy$ (right) from Ref.~\cite{Chluba2013greens}.~\notebooklink{notebooks/paper\_figures/mu\_y\_vs\_injection\_redshift.ipynb}}
\label{fig:mu_y_vs_zh}
\end{figure}

Figure~\ref{fig:mu_y_vs_zh} shows the distortion amplitudes $\mu/(\drho)$ and $4y/\drho$ as a function of injection redshift for single-burst injections at 40 redshifts spanning $z_h =\,$$10^3$ to $3 \times 10^6$, where $\mu$ and $y$ are extracted from each PDE spectrum via the decomposition of Appendix~\ref{app:decomposition} with the temperature shift removed. 
In the $\mu$-era ($z_h \gtrsim\,$$3 \times 10^5$), the PDE recovers the analytic limit $\mu/\drho = 3/\kc \approx 1.401$ (Eq.~\eqref{eq:mu_drho}) to within a few percent, with thermalization progressively suppressing $\mu$ at higher redshifts.
In the $y$-era ($z_h \lesssim\,$$10^4$), the PDE reproduces $4y/\drho = 1$ (Sec.~\ref{sec:thermalization_regimes}) to within $1.2\%$, and follows $\Jy$ to within $0.6\%$.
The solid orange lines show the analytic predictions from the visibility functions $\Jbb(z)\,\Jmu(z)$ (left) and $\Jy(z)$ (right) of Ref.~\cite{Chluba2013greens}, which encode the thermalization efficiency at each redshift (Sec.~\ref{sec:greens}).

\paragraph{Visibility functions.}
Refs.~\cite{Chluba2013greens,Chluba:2013kua,Chluba2015photon} showed that the three visibility functions introduced in Eq.~\eqref{eq:gth} can be parameterized as
\begin{align}
    \Jbb(z) &= A\,e^{-(z/z_\mathrm{th})^{\alpha_\mathrm{th}}}\bigl[1 - B\,(z/z_\mathrm{th})^{\beta}\bigr]\,, \label{eq:jbb} \\
    \Jmu(z) &= 1 - \exp\!\bigl[-\bigl((1+z)/z_\mu\bigr)^{\alpha_\mu}\bigr]\,, \label{eq:jmu} \\
    \Jy(z)  &= \bigl[1 + \bigl((1+z)/z_y\bigr)^{\alpha_y}\bigr]^{-1}\,. \label{eq:jy}
\end{align}
The thermalization redshift $z_\mathrm{th} \approx 1.98 \times 10^6$ and exponent $\alpha_\mathrm{th} = 5/2$ are analytically derived from the redshift scaling of the double Compton opacity~\cite{Hu1993} and are not fit parameters.
The remaining seven parameters $(A, B, \beta, z_\mu, \alpha_\mu, z_y, \alpha_y)$ govern the branching: $\Jbb$ measures thermalization efficiency (exponentially suppressed at $z \gg z_\mathrm{th}$, approaching unity at $z \ll z_\mathrm{th}$); $\Jmu$ transitions from $0$ in the $y$-era ($z \lesssim\,$$10^4$) to $1$ in the $\mu$-era; and $\Jy$ approaches unity at low $z$ and zero at high $z$.
Literature values for these parameters~\cite{Chluba2013greens,Chluba:2013kua,Chluba2015photon} appear in the ``Literature'' column of Table~\ref{tab:visibility_params}.

We verify that the PDE solver captures the correct thermalization physics by independently deriving the visibility function parameters (Eqs.~\eqref{eq:jbb}--\eqref{eq:jy}) from PDE spectra.
At each of 118 injection redshifts spanning $z_h = 3\times 10^3$ to $5\times 10^6$, we run a single-burst PDE simulation at $n_\mathrm{grid} = 4000$ frequency points and subtract the unobservable temperature shift as described in Sec.~\ref{sec:thermalization_regimes}.
We fix the analytically derived parameters $z_\mathrm{th} = 1.98 \times 10^6$ and $\alpha_\mathrm{th} = 5/2$ and perform a global fit of the remaining seven parameters in the visibility function Ansatz (Eqs.~\eqref{eq:jbb}--\eqref{eq:jy}) to the full set of PDE spectra simultaneously, minimizing the $x^3$-weighted spectral residual between the model~\eqref{eq:gth} and the number-conserving-stripped PDE across all redshifts and frequencies $x \in [0.5, 20]$.
Before fitting, we subtract the adiabatic-cooling distortion that accumulates in every run, measured from an otherwise identical no-injection run; it otherwise biases the extracted visibility at the high redshifts where thermalization has suppressed the injected signal.

\begin{figure}[!t]
\centering
\includegraphics[width=\textwidth]{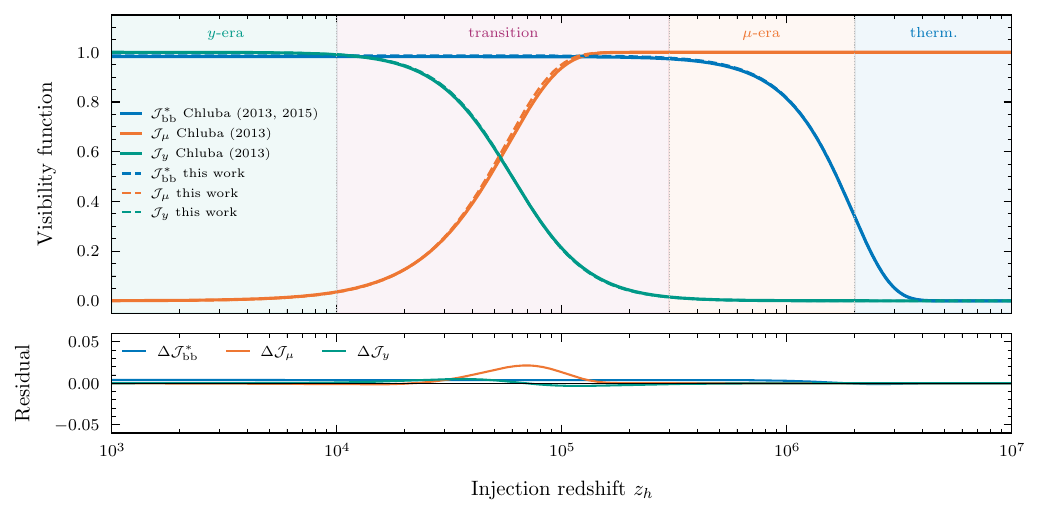}
\caption{Visibility functions derived from PDE spectra.
\emph{Top:} Solid lines show the literature fitting formulas from Refs.~\cite{Chluba2013greens,Chluba2015photon}. Dashed lines show the visibility fitted to 118 PDE spectra ($n_\mathrm{grid}=4000$).
\emph{Bottom:} Absolute difference between PDE-derived and literature visibility functions. Deviations remain below 0.022 across all redshifts.
~\notebooklink{notebooks/paper\_figures/visibility\_functions.ipynb}}
\label{fig:visibility}
\end{figure}

As shown in Fig.~\ref{fig:visibility}, the PDE-derived visibility functions are in good agreement with the literature fitting formulas of Refs.~\cite{Chluba2013greens,Chluba:2013kua,Chluba2015photon}.
Table~\ref{tab:visibility_params} lists the fitted parameters alongside the literature values.
The five primary parameters are recovered to within 3\% of the values from Refs.~\mbox{\cite{Chluba2013greens,Chluba2015photon}}.
The two correction parameters ($B$, $\beta$) in $\Jbb$ are less well constrained because their correction term $1 - B(z/z_\mathrm{th})^\beta$ only deviates appreciably from unity at $z \gtrsim 2\times 10^6$, where thermalization has already suppressed the distortion signal below detectability.
The fitted $\Jbb$ also agrees with the direct chemical-potential extraction to $0.9\%$ rms over $3\times10^5 \le z_h \le 3\times10^6$.

\begin{table}[t]
\centering
\begin{tabular}{llccr}
\hline
Function & Parameter & PDE fit & Literature & $\Delta$ (\%) \\
\hline
$\Jy$ & $z_y$ & $60\,100$ & $60\,000$ & $+0.2$ \\
 & $\alpha_y$ & $2.63$ & $2.58$ & $+1.9$ \\[2pt]
$\Jmu$ & $z_\mu$ & $56\,400$ & $58\,000$ & $-2.7$ \\
 & $\alpha_\mu$ & $1.93$ & $1.88$ & $+2.9$ \\[2pt]
$\Jbb$ & $A$ & $0.987$ & $0.983$ & $+0.4$ \\
 & $B$ & $0.043$ & $0.038$ & $+13$ \\
 & $\beta$ & $2.40$ & $2.29$ & $+4.9$ \\
 & $z_\mathrm{th}$ & \multicolumn{2}{c}{$1.98\times 10^6$ (fixed)} & --- \\
 & $\alpha_\mathrm{th}$ & \multicolumn{2}{c}{$5/2$ (fixed)} & --- \\
\hline
\end{tabular}
\caption{Visibility function parameters derived from 118 \spectroxide{}-produced single-burst energy injections ($n_\mathrm{grid}=4000$), compared with literature values from Refs.~\cite{Chluba2013greens,Chluba2015photon}. All seven parameters come from a single $x^3$-weighted spectral fit over $x \in [0.5, 20]$ with the adiabatic-cooling baseline subtracted. The thermalization redshift $z_\mathrm{th}$ and exponent $\alpha_\mathrm{th} = 5/2$ are analytically derived and held fixed (see text).}
\label{tab:visibility_params}
\end{table}

\begin{figure}[!t]
\centering
\includegraphics[width=\textwidth]{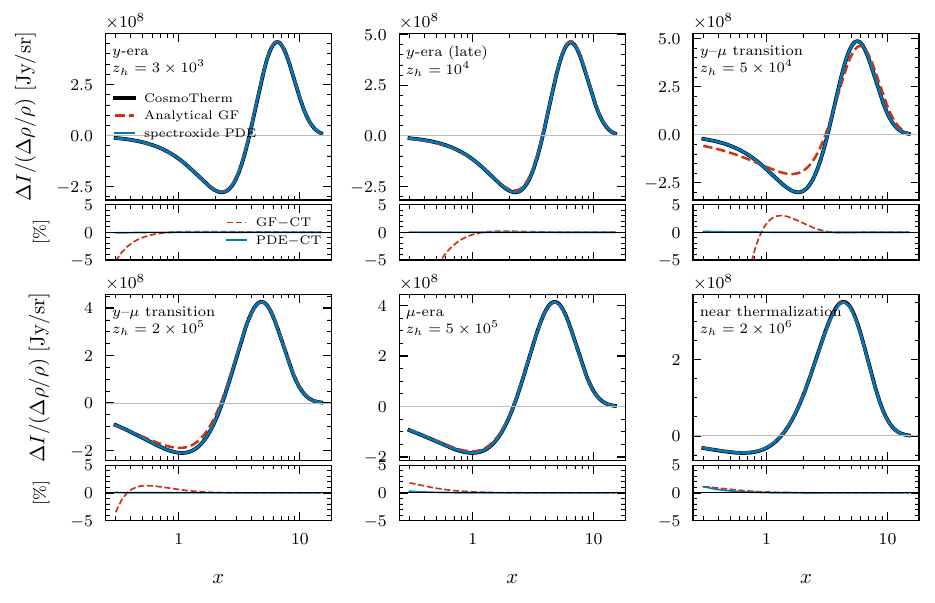}
\caption{Intensity distortion $\Delta I_\nu$ per unit $\drho$ from single-burst energy injection at six representative redshifts spanning the $y$-era through thermalization.
Each panel shows the \spectroxide{} PDE solver (blue), analytical Green's function (red dashed), and \cosmotherm{}~\cite{Chluba2012cosmotherm,Chluba2013greens} (black); the unobservable temperature-shift component $(\Delta T/T)\,\Gbb(x)$ has not been subtracted from any curve, in order to show the growing $\Gbb$ component at high $z_h$ as thermalization converts injected energy into a blackbody temperature correction (contrast with Fig.~\ref{fig:dm_comparison}, where the temperature shift is subtracted to isolate the observable distortion). We also restore the $\Gbb$ component to the \cosmotherm{} Green's function entries.
At low $z_h$ the distortion has the characteristic $y$-shape; at high $z_h$ Compton scattering redistributes the energy into a Bose--Einstein distribution with chemical potential $\mu$, while the temperature-shift ($\Gbb$) component grows as thermalization converts more of the injected energy into a blackbody temperature correction.
Bottom panels show the peak-normalized residual relative to \cosmotherm{}~\cite{Chluba2012cosmotherm}.
~\notebooklink{notebooks/paper\_figures/cosmotherm\_comparison.ipynb}}
\label{fig:spectral_shapes}
\end{figure}

\paragraph{Spectral comparison with \cosmotherm{}.}
Beyond the integrated $\mu$ and $y$ parameters, we can test whether the full frequency-dependent spectral shape of individual Green's function entries is correct.
We compare against the publicly available Green's function table from \cosmotherm{} v1.0.3~\cite{Chluba2012cosmotherm,Chluba2013greens}.
We fix \spectroxide{} to use the cosmological parameters provided in the \cosmotherm{} v1.0.3 documentation to generate these Green's functions ($h = 0.71$, $\Omega_\mathrm{b} = 0.044$, $\Omega_\mathrm{m} = 0.26$, $Y_p = 0.24$, $T_\mathrm{CMB} = \qty{2.726}{\kelvin}$).%
\footnote{We would like to thank Jens Chluba for alerting us to a possible discrepancy between the values used here and those used to generate the tabulated Green's functions from \cosmotherm{}. We have checked that the Green's functions are largely insensitive to small shifts in $\Omega_\mathrm{m}$.}
The intensity distortion shown in Fig.~\ref{fig:spectral_shapes} and subsequent figures is related to the occupation-number distortion by $\Delta I_\nu = 4\pi\nu^3\,\dn(x)$.
Figure~\ref{fig:spectral_shapes} compares the \spectroxide{} PDE distortion with entries from the publicly available \cosmotherm{} Green's function table at six representative injection redshifts spanning the $y$-era through the onset of thermalization.
The residual panels show sub-percent agreement in spectral shape over much of the range of $x$ that we show. Over the plotted range, the rms PDE residual is at most $0.1\%$ of the peak for every $z_h$. 

In Fig.~\ref{fig:spectral_shapes}, we also compare the \spectroxide{} results with the analytic Green's function (Eq.~\eqref{eq:gth}) (red dashed). These show larger deviations in the $\mu$--$y$ transition, where the three-component Ansatz is an inherently limited approximation.
However, deep in the $y$, $\mu$, and thermalization eras, the analytic formulas agree well with the \spectroxide{} results. 

\begin{figure}[!t]
\centering
\includegraphics[width=\textwidth]{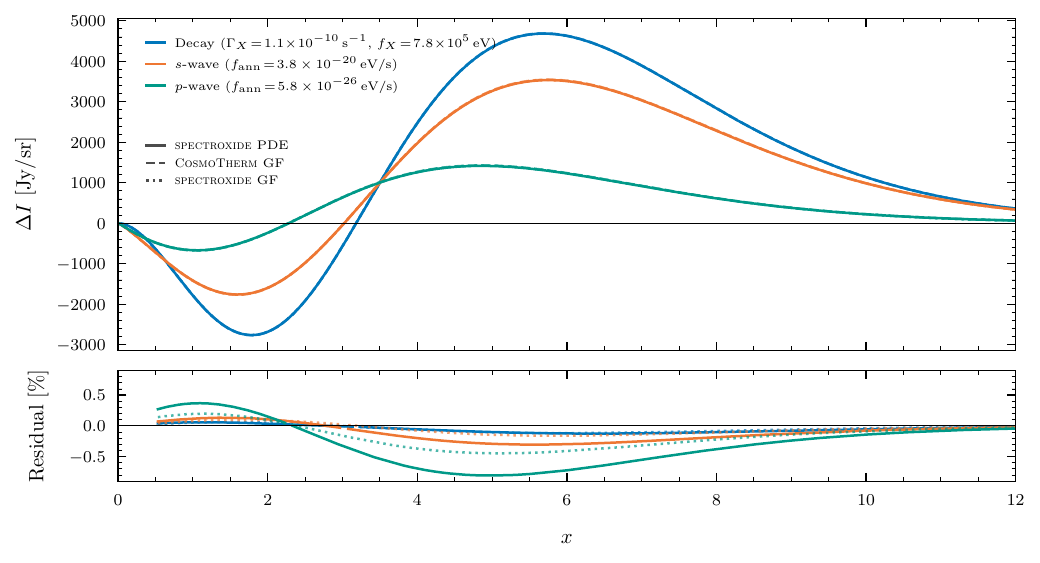}
\caption{Spectral distortions $\Delta I$ [Jy/sr] for three dark matter scenarios, each with $\drho \sim 10^{-5}$: \spectroxide{} PDE (solid colored) compared against \cosmotherm{} Green's function convolution (dashed) and \spectroxide{} PDE-derived Green's function convolution (dotted).
Scenarios: decaying particle with $\Gamma_X = \qty{1.1e-10}{\per\second}$ and $f_X = \qty{7.8e5}{\eV}$ (blue); $s$-wave annihilation with $f_\mathrm{ann} = \qty{3.8e-20}{\eV\per\second}$ (orange); $p$-wave annihilation with $f_\mathrm{ann} = \qty{5.8e-26}{\eV\per\second}$ (green).
All spectra have the unobservable temperature shift subtracted by enforcing $\int x^2 \dn\,dx = 0$.
Bottom panel shows the residual vs.\ \cosmotherm{} for each scenario; agreement is $\lesssim 2\%$ across the FIRAS/FOSSIL band.
~\notebooklink{notebooks/paper\_figures/dm\_scenario\_comparison.ipynb}}
\label{fig:dm_comparison}
\end{figure}

\paragraph{Dark matter scenarios.}
The single-burst comparisons above test each injection redshift independently, but realistic scenarios inject energy continuously over a range of redshifts. Our code must therefore be robust against errors that accumulate during extended integration for such scenarios.
Figure~\ref{fig:dm_comparison} shows the spectral distortion for three dark matter scenarios---a decaying particle ($\Gamma_X = \qty{1.1e-10}{\per\second}$, $f_X = \qty{7.8e5}{\eV}$), $s$-wave DM annihilation ($f_\mathrm{ann} = \qty{3.8e-20}{\eV\per\second}$), and $p$-wave annihilation ($f_\mathrm{ann} = \qty{5.8e-26}{\eV\per\second}$)---each with $\drho \sim 10^{-5}$, computed with the \spectroxide{} PDE solver.
As an independent cross-check, we convolve the same injection histories with precomputed Green's function tables via Eq.~\eqref{eq:convolution}: one table from \cosmotherm{} (dashed lines) and one constructed from \spectroxide{} single-burst PDE runs at many redshifts (dotted lines).
The \spectroxide{} PDE agrees with both the \cosmotherm{} and \spectroxide{} Green's function convolutions to $\lesssim 2\%$ for all three scenarios (Fig.~\ref{fig:dm_comparison}, bottom panel).

\begin{figure}[!t]
\centering
\includegraphics[width=\textwidth]{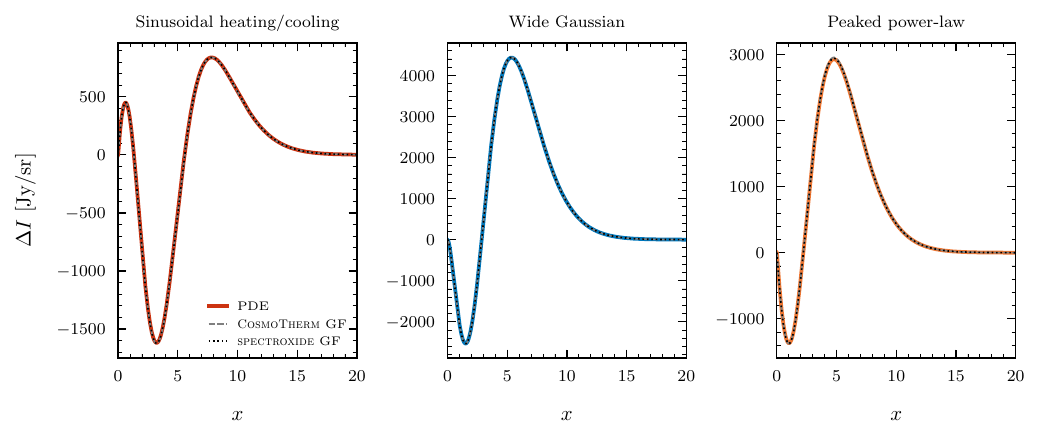}
\caption{Stress tests with pathological injection histories: sinusoidal heating/cooling (left), wide Gaussian spanning the $\mu$--$y$ transition (center), and peaked power-law peaking at $z = 9 \times 10^5$ (right).
Each panel compares the \spectroxide{} PDE (solid), \cosmotherm{} Green's function convolution (dashed), and \spectroxide{} Green's function table convolution (dotted).
~\notebooklink{notebooks/paper\_figures/pathological\_heating.ipynb}}
\label{fig:pathological}
\end{figure}

\paragraph{Stress tests.}
The tests above validate the solver for physically motivated scenarios, but they do not probe whether the solver is robust to arbitrary injection profiles.
We stress-test this by running the PDE solver with three pathological custom injection histories, and independently convolving the same heating rates with our PDE-derived Green's function table (via Eq.~\eqref{eq:convolution}) and the \cosmotherm{} Green's function table:
\begin{enumerate}
    \item \emph{Sinusoidal heating/cooling}: $d(\drho)/dz = A_\mathrm{sin}\,\sin[k(1+z)]$ with $A_\mathrm{sin} = 5 \times 10^{-10}$ and $k = 2\pi / (5 \times 10^4)$, active over $z \in [10^3, 5 \times 10^5]$, where positive and negative contributions must cancel precisely.
    \item \emph{Wide Gaussian}: $d(\drho)/dz = A_\mathrm{G}\,\exp[-(z - z_0)^2 / 2\sigma^2]$ with $A_\mathrm{G} = 10^{-5}/(\sigma\sqrt{2\pi})$, $z_0 = 8 \times 10^4$, and $\sigma = 4 \times 10^4$, spanning the $\mu$-to-$y$ transition where the GF shape approximation is least accurate.
    \item \emph{Peaked power-law}: $d(\drho)/dz = A_\mathrm{pl}\,z^3\,\exp(-z/z_c)$ with $A_\mathrm{pl} = 10^{-5}/(6\,z_c^4)$ and $z_c = 3 \times 10^5$, peaking at $z = 3z_c = 9 \times 10^5$ to exercise the thermalization regime.
\end{enumerate}
Each amplitude is chosen to give $\drho \sim 10^{-5}$.
In all cases, the PDE solver produces stable, well-converged spectra.
Both the \cosmotherm{} and \spectroxide{} Green's function table convolutions agree with the continuous PDE to sub-percent RMS in spectral shape over \qtyrange{30}{800}{\GHz} (Fig.~\ref{fig:pathological}).
These tests provide a stringent internal consistency check on the PDE solver: by independently convolving the same injection histories with the Green's function, we verify that the solver produces the correct spectral shapes across sign-changing, broad, and sharply peaked injection profiles spanning the $\mu$-era, $y$-era, and thermalization regime.
The sub-percent agreement with \cosmotherm{} further confirms consistency with published results.
Because all three histories are run through the PDE and, independently, through two separately constructed Green's function tables, this is among the most stringent end-to-end consistency tests in the paper.

\subsection{Photon injection validation}
\label{sec:photon_validation}

Unlike heat injection, where the \cosmotherm{} Green's function table provides a comprehensive independent benchmark, validation of photon injection is more limited: no open-source PDE-level comparison code exists.
We therefore validate the photon injection solver using three complementary checks: comparison with the photon Green's functions (Eq.~\eqref{eq:photon_gf_mu}), FIRAS constraints on monochromatic photon injection following the methodology of Ref.~\cite{Chluba2015photon}, and dark photon oscillations as an end-to-end application (Sec.~\ref{sec:dark_photon}).
We additionally compared \spectroxide{} monochromatic photon injection results against published plots produced with \cosmotherm{} from Ref.~\cite{Chluba2015photon} and found good agreement, although a quantitative analysis is difficult to perform in this case.

\begin{figure}[!t]
\centering
\includegraphics[width=\textwidth]{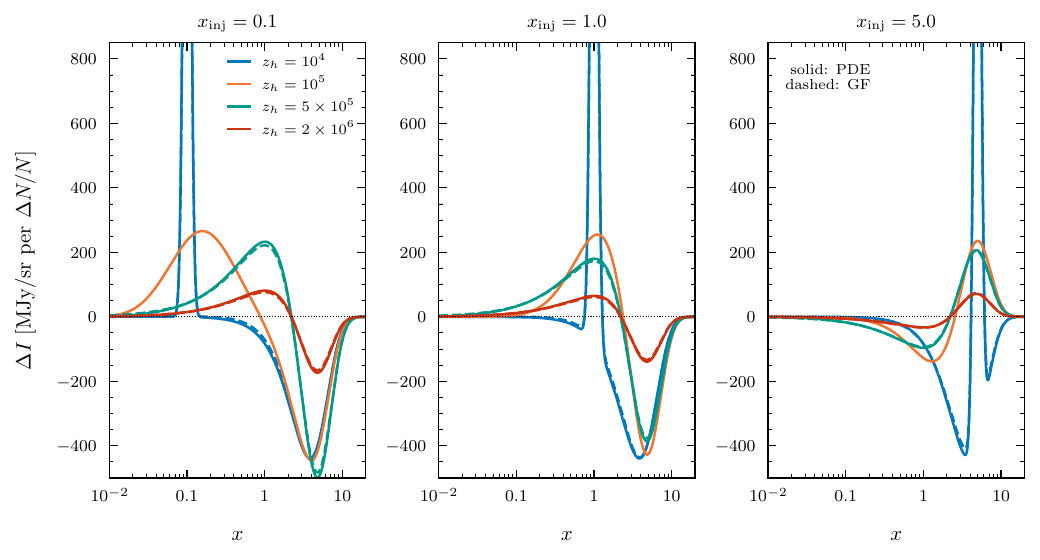}
\caption{Spectral distortions from monochromatic photon injection at three injection frequencies $x_\mathrm{inj} = 0.1$, $1$, and $5$ (panels left to right), each computed at four injection redshifts spanning the $y$-era ($z_h = 10^4$) through near-thermalization ($z_h = 2\times10^6$).
Solid lines show the PDE solver result; dashed lines show the analytic Green's function approximation for redshifts where the pure $\mu$ or $y$ regime applies (GF omitted at $z_h = 10^5$, which lies in the intermediate transition region).
At low injection redshifts, a sharp spectral feature persists near $x_\mathrm{inj}$ because Compton scattering is too slow to redistribute it; at higher redshifts, the feature is progressively broadened and the distortion approaches the smooth $\mu$/$y$ shape expected from heat injection.
The $y$-era GF uses the numerical $\tau_{\mathrm{ff}}$ survival probability (Eq.~\eqref{eq:tau_ff}); the $\mu$-era GF uses the analytic $x_c/x_\mathrm{inj}$ formula (Eq.~\eqref{eq:ps}).
Amplitudes are per unit $\Delta N_\gamma / N_\gamma$.
~\notebooklink{notebooks/paper\_figures/photon\_injection\_spectra.ipynb}}
\label{fig:photon_injection}
\end{figure}

Figure~\ref{fig:photon_injection} shows PDE solutions for monochromatic photon injection at three injection frequencies ($x_\mathrm{inj} = 0.1$, $1$, and $5$) and four injection redshifts spanning the $y$-era through near-thermalization.
At all three frequencies, a spectral bump persists near $x_\mathrm{inj}$ for low injection redshifts ($z_h = 10^4$), where Compton scattering is too slow to redistribute the injected photons away from their injection frequency.
At higher injection redshifts, Compton scattering is more efficient, which progressively broadens and shifts the bump.
Dashed lines show the analytic Green's function approximation of Ref.~\cite{Chluba2015photon} (Sec.~\ref{sec:greens}) for redshifts solidly in the $y$-era ($z_h = 10^4$) or $\mu$-era ($z_h \geq 5\times10^5$).
We omit the GF comparison at $z_h = 10^5$, which falls in the intermediate $\mu$--$y$ transition regime where there is no analytic GF.
Differences between the PDE and the dashed curves in Fig.~\ref{fig:photon_injection} are due to the analytic approximation breaking down rather than solver error.

\begin{figure}[!t]
\centering
\includegraphics[width=0.6\textwidth]{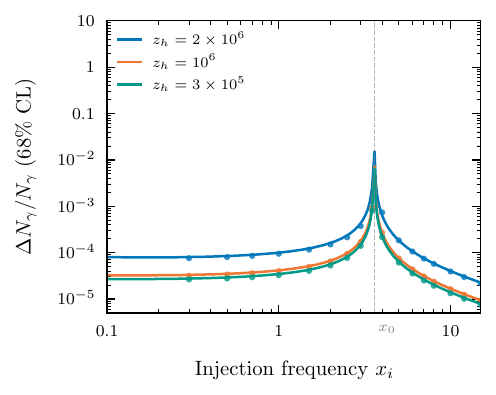}
\caption{COBE/FIRAS 68\% CL upper limits on monochromatic photon injection $\Delta N_\gamma / N_\gamma$ as a function of injection frequency $x_i$, at three injection redshifts.
The sharp peak near $x_0 \approx 3.60$ marks the balanced frequency where $\mu$ vanishes (Sec.~\ref{sec:photon_gf}).
The solid line is computed from the $\mu$-era photon Green's function (Eq.~\eqref{eq:photon_gf_mu}) and validates the PDE solver.
~\notebooklink{notebooks/paper\_figures/firas\_photon\_limits.ipynb}}
\label{fig:firas_photon}
\end{figure}

As an additional check, we also derive COBE/FIRAS constraints on monochromatic photon injection, following the methodology of Ref.~\cite{Chluba2015photon}.
For each injection frequency $x_i$ and redshift $z_h$, we run the PDE solver, extract the $\mu$ and $y$ distortions per unit $\Delta N_\gamma / N_\gamma$, and invert the FIRAS 68\% CL limits ($|\mu| < 4.5 \times 10^{-5}$, $|y| < 7.5 \times 10^{-6}$), keeping the tighter of the two, to obtain the maximum allowed photon injection.
Figure~\ref{fig:firas_photon} shows the PDE-derived constraints, with the constraint obtained using the analytic $\mu$-era Green's function (Eq.~\eqref{eq:photon_gf_mu}) overlaid for comparison.

At low injection frequencies ($x_c \ll x_i \ll x_0$), the injected photons survive DC/BR absorption ($P_s \approx 1$). They add photon number but little energy, so each photon produces a negative $\mu$ of nearly fixed size (Eq.~\eqref{eq:photon_gf_mu}), and the constraint is roughly independent of $x_i$.
As $x_i$ approaches the balanced frequency $x_0 \approx 3.60$ (Sec.~\ref{sec:photon_gf}), the $\mu$-distortion per photon vanishes and the constraint weakens sharply, producing the characteristic peak in allowed $\Delta N_\gamma / N_\gamma$.
Above $x_0$, injected photons carry more energy than number and produce positive $\mu$, so the constraint tightens again.
At higher injection redshifts, DC/BR thermalization is more efficient and the overall constraint is weaker because a larger fraction of the injected energy is absorbed into the blackbody.

\section{Dark photon oscillations}
\label{sec:dark_photon}
Having validated the solver for both heat and photon injection, we now apply it to a concrete new-physics scenario.
Dark photon oscillations provide a good test case because they combine photon depletion (a frequency-dependent source term) with thermalization physics across multiple eras, exercising the full solver machinery in a single application.

A dark photon $A'$ is a massive hidden $U(1)$ gauge boson that kinetically mixes with the Standard Model photon via the Lagrangian term $-(\epsilon/2)\,F_{\mu\nu}F'^{\mu\nu}$, where $\epsilon$ is the kinetic mixing parameter~\cite{Holdom1986,Mirizzi2009}.
In the early universe, the photon propagates through a plasma and acquires an effective mass equal to the plasma frequency, $m_\gamma^2 \simeq \omega_\mathrm{pl}^2 = 4\pi\alpha\,n_e / m_e$.
Oscillations between the photon and dark photon mass eigenstates are resonantly enhanced when $\omega_\mathrm{pl}(z) = m_{A'}$, analogous to the MSW resonance for neutrinos~\cite{Caputo2020,Caputo2020modeling,Chluba2024darkphoton}.
Since $\omega_\mathrm{pl} \propto \sqrt{n_e}$ decreases monotonically with the expansion before recombination, each dark photon mass maps to a unique resonance redshift $z_\mathrm{res}$: masses $m_{A'} \sim \qty{e-5}{\eV}$ experience a resonant conversion at $z_\mathrm{res} \sim 7 \times 10^5$ (in the $\mu$-era), while masses $m_{A'} \lesssim \qty{e-10}{\eV}$ undergo resonant conversions after recombination ($z_\mathrm{res} \lesssim 10^3$)~\cite{Mirizzi2009,Caputo2020}. After recombination, neutral hydrogen also contributes to the photon mass, $m_\gamma^2 \simeq \omega_\mathrm{pl}^2 - 4\pi\alpha_\mathrm{H}\,n_\mathrm{HI}\,\omega^2$ with polarizability $\alpha_\mathrm{H} = 4.5\,a_0^3$ and $a_0$ the Bohr radius, which makes the resonance redshift depend on photon frequency~\mbox{\cite{Caputo2020,Caputo2020modeling}}. Here, we neglect this term, as Ref.~\mbox{\cite{Chluba2024darkphoton}} does, so that our limits compare directly with theirs; this effect is only important at low $m_{A'}$ where the resonance occurs after recombination, and we refer the reader to Ref.~\mbox{\cite{Caputo2020}} for a more careful treatment including inhomogeneities.
For all the cases we consider in this work, the plasma is highly uniform and spatial inhomogeneities in $\omega_\mathrm{pl}$ are negligible~\cite{Caputo2020modeling}; this would not hold for very late resonances at $z \lesssim 200$, where baryon density fluctuations can broaden the resonance~\cite{Caputo2020}.

The resonance is extremely narrow and is well described by the narrow-width approximation~\cite{Mirizzi2009,Caputo2020,Caputo2020modeling}.
The probability that a photon of dimensionless frequency $x$ converts into a dark photon is $p(x) = 1 - e^{-\gamma_\mathrm{con}/x}$, where $\gamma_\mathrm{con}$ is the dimensionless conversion parameter~\cite{Chluba2024darkphoton}
\begin{equation}
    \gamma_\mathrm{con} = \frac{\pi\epsilon^2m_{A'}^2}{\Tz(z_\mathrm{res})\,H(z_\mathrm{res})(1+z_\mathrm{res})}\left|\frac{d\ln\omega_\mathrm{pl}^2}{dz}\right|_{z_\mathrm{res}}^{-1}\,.    \label{eq:gamma_con}
\end{equation}
The initial photon depletion imposed at $z_\mathrm{res}$ is
\begin{equation}
    \dn(x, z_\mathrm{res}) = -p(x)\,\npl(x) = -\bigl[1 - e^{-\gamma_\mathrm{con}/x}\bigr]\,\npl(x)\,.    \label{eq:dp_depletion}
\end{equation}
This serves as an initial condition for the subsequent evolution; the observed spectral shape at $z=0$ differs once Compton scattering and DC/BR reprocess the depletion (see below).
The spectral shape of the distortion depends on $z_\mathrm{res}$.
For pre-recombination resonances ($z_\mathrm{res} \gtrsim 1100$), the photon depletion is reprocessed by Compton scattering and DC/BR emission and the interplay between these processes determines the final spectral shape.
For post-recombination resonances ($z_\mathrm{res} \lesssim 1100$), Compton scattering freezes out and the distortion is locked in at the injection shape $\propto \npl(x)/x$~\cite{Arsenadze2025}.

\paragraph{FIRAS constraints.}
As a demonstration of \spectroxide{}'s capabilities (rather than a new physical result), we reproduce the 95\% CL upper limits on $\epsilon$ as a function of $m_{A'}$, following the methodology of Refs.~\cite{Chluba2024darkphoton,Arsenadze2025}.

For each dark photon mass $m_{A'}$, we run the PDE solver at a reference mixing parameter $\epsilon_\mathrm{ref}$ to obtain the reprocessed spectral distortion $\dn_\mathrm{ref}(x)$.
In the linear regime ($\gamma_\mathrm{con} \ll 1$), the distortion scales as $\dn \propto \gamma_\mathrm{con} \propto \epsilon^2$ (Eq.~\eqref{eq:gamma_con}), so we define a mass-dependent spectral template $\mathcal{T}(x;\,m_{A'}) \equiv \dn_\mathrm{ref} / \gamma_\mathrm{con,ref}$ that captures the reprocessed shape independently of $\epsilon$.
This allows the FIRAS fit to scan over $\epsilon$ without re-running the PDE: the model prediction at any mixing parameter is $\dn_\mathrm{model} = \gamma_\mathrm{con}(\epsilon) \times \mathcal{T}(x;\,m_{A'})$.
We fit the model
\begin{equation}
  I_\mathrm{obs}(\nu) - B(\nu, T) = 4\pi \nu^3\, \gamma_\mathrm{con}\, \mathcal{T}\bigl(x(T);\,m_{A'}\bigr) + G_0\, \nu^2 B(\nu, T_d)
  \label{eq:dp_firas_model}
\end{equation}
to the 43 FIRAS monopole residuals using the full $43 \times 43$ frequency-frequency covariance matrix~\cite{Fixsen1996}. The first term on the right-hand side converts the occupation-number template $\mathcal{T}$ to specific intensity. The second term is a galactic dust foreground following Ref.~\cite{Fixsen1996}. Residual galactic emission is not perfectly subtracted from the FIRAS monopole and is partially degenerate with broadband distortion shapes, so we marginalize over a fixed-shape $\nu^2 B(\nu, T_d)$ template with $T_d = \qty{9}{\K}$ and free amplitude $G_0$.
Here, $B(\nu, T) = (4\pi\nu^3)/(e^{2\pi\nu/T} - 1)$ is the Planck function.
Because the CMB reference temperature $T$ is itself a free parameter (encoding the unobservable temperature shift), both the residuals and the template shape $\mathcal{T}(x(T))$ with $x(T) = 2\pi\nu / T$ depend nonlinearly on $T$~\cite{Arsenadze2025}.
We therefore profile over $T$ by scanning a grid of trial temperatures around $T_0$. Then, at each $T$, the best-fit $\gamma_\mathrm{con}$ and $G_0$ are obtained analytically (the model is linear in both), and we take the $T$ that minimizes $\chi^2$. The Gaussian log-likelihood corresponding to Eq.~\eqref{eq:dp_firas_model} is
\begin{equation}
  -2\log L(\epsilon^2;\,T,\,G_0) = \sum_{i,j=1}^{43} \Delta_i\,(C^{-1})_{ij}\,\Delta_j \,,
  \label{eq:dp_firas_loglike}
\end{equation}
where $C$ is the FIRAS covariance and $\Delta_i \equiv I_\mathrm{obs}(\nu_i) - B(\nu_i,T) - 4\pi\nu_i^3\,\gamma_\mathrm{con}\,\mathcal{T}\bigl(x_i(T);\,m_{A'}\bigr) - G_0\,\nu_i^2\,B(\nu_i,T_d)$ is the residual at frequency $\nu_i$ (i.e., LHS minus RHS of Eq.~\eqref{eq:dp_firas_model}). Following Ref.~\cite{Arsenadze2025}, we set upper limits on $\epsilon^2$ at fixed $m_{A'}$ from the profile likelihood ratio
\begin{equation}
  \lambda(\epsilon^2) \equiv -2\log\frac{L\bigl(\epsilon^2;\,\hat{\hat T},\hat{\hat G}_0\bigr)}{L\bigl(\hat{\epsilon}^2;\,\hat{T},\hat{G}_0\bigr)}\,,
  \label{eq:dp_firas_ts}
\end{equation}
where $\hat{\hat T}$ and $\hat{\hat G}_0$ maximize $L$ at the specified $\epsilon^2$, while $\hat{\epsilon}^2$, $\hat T$, $\hat G_0$ maximize $L$ globally. By Wilks' theorem, $\lambda$ asymptotically follows a $\chi^2_1$ distribution under the null hypothesis given by the specified value of $\epsilon^2$, so the one-sided 95\% CL upper limit corresponds to $\lambda = 2.71$.

In \spectroxide{}, computing the dark photon template and deriving FIRAS constraints requires only a few lines:
\begin{lstlisting}
import numpy as np
from spectroxide import solve
from spectroxide.dark_photon import gamma_con
from spectroxide.firas import FIRASData

# 1. NWA gamma_con at (eps_ref, m_Aprime) for template rescaling
eps_ref, m = 1e-9, 1e-7  # eV
gc_ref, _ = gamma_con(eps_ref, m)

# 2. PDE with dark photon IC applied automatically at z_res
res = solve(injection={"type": "dark_photon_resonance",
                       "epsilon": eps_ref, "m_ev": m},
    z_end=100, n_points=4000,
)
x = np.asarray(res.x)
dn_per_gc = np.asarray(res.delta_n) / gc_ref

# 3. Floating-T profile likelihood vs FIRAS
firas = FIRASData()
fit = firas.profile_limit_floating_T(
    lambda xx: np.interp(xx, x, dn_per_gc),
    cl=0.95,
)
gc_95 = fit["upper_limit"]
eps_95 = eps_ref * np.sqrt(gc_95 / gc_ref)
\end{lstlisting}

\begin{figure}[!t]
\centering
\includegraphics[width=\textwidth]{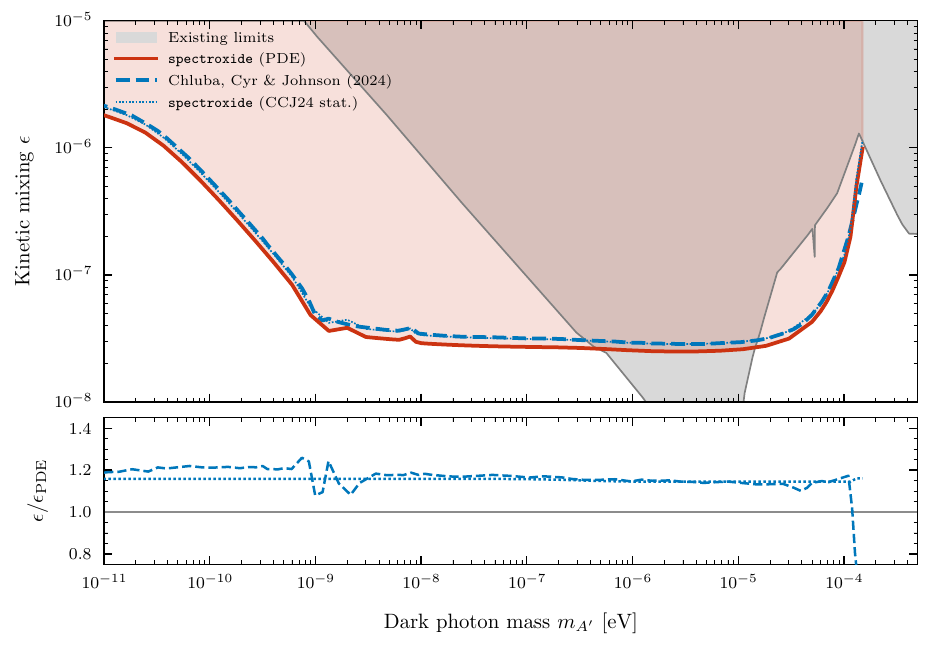}
\caption{FIRAS 95\% CL upper limits on dark photon kinetic mixing $\epsilon$ as a function of mass $m_{A'}$, derived from PDE solver spectral templates using profile likelihood with floating blackbody temperature~\cite{Arsenadze2025} and the full $43 \times 43$ FIRAS covariance matrix, marginalized over a galactic dust isotropic component $\propto \nu^2 B(\nu, \qty{9}{\K})$ following Ref.~\cite{Fixsen1996}.
The red shaded region is excluded.
The dashed curve shows the published limits of Ref.~\cite{Chluba2024darkphoton}; the dotted curve applies their statistic (diagonal errors, $\Delta\chi^2 = 3.84$) to the same \spectroxide{} PDE  templates. The lower panel shows both relative to our fiducial \spectroxide{} limits.
Gray shaded regions show existing laboratory constraints from Cavendish--Coulomb tests~\cite{Tu2005}, light-shining-through-wall experiments (ALPS~\cite{ALPS2010}, CROWS~\cite{CROWS2013}), and the Dark SRF pathfinder~\cite{DarkSRF2023}, compiled from Ref.~\cite{AxionLimits}.
Constraints are strongest ($\epsilon \lesssim 2.5 \times 10^{-8}$) in the transition region ($m_{A'} \sim \qty{e-6}{\eV}$).
~\notebooklink{notebooks/paper\_figures/dark\_photon\_constraints.ipynb}}
\label{fig:dp_firas}
\end{figure}

Figure~\ref{fig:dp_firas} shows the resulting exclusion region from the PDE solver at 70 representative masses spanning \qtyrange{e-12}{1.5e-4}{\eV}, alongside laboratory constraints for comparison.
The strongest constraints ($\epsilon \lesssim 2.5 \times 10^{-8}$, corresponding to $\gamma_\mathrm{con} \sim 6 \times 10^{-5}$) occur in the $\mu$--$y$ transition region ($m_{A'} \sim \qty{e-6}{\eV}$) and are well within the linear regime.

We also overlay the COBE/FIRAS constraints from Ref.~\cite{Chluba2024darkphoton}, and observe a slight disagreement between the two results due to our statistical methodology. To verify that the visible offset between the two results is statistical methodology rather than solver physics, we re-derived limits from the same PDE templates using the statistic of Ref.~\cite{Chluba2024darkphoton}, which reproduces their published curve~\mbox{\cite{AxionLimits}} well, as shown in Fig.~\ref{fig:dp_firas}. The discrepancy arises from the use of a two-sided CL to set upper limits on $\epsilon$ in Ref.~\mbox{\cite{Chluba2024darkphoton}}.

At high masses, the constraint weakens as DC/BR thermalization progressively erases the distortion ($\Jbb \to 0$).
We terminate the constraint around $m_{A'} \sim \qty{1.5e-4}{\eV}$, where the conversion probability reaches $\gamma_\mathrm{con} \approx 0.1$, depleting ${\sim}10\%$ of CMB photons at $x \sim 1$ and entering the strong distortion regime outside the linear approximation used here~\cite{Chluba2024darkphoton}. Additionally, we note that for $m_{A'}\gtrsim \qty{2e-4}{\eV}$, where the kinetic mixing parameter is constrained to $\epsilon \gtrsim 10^{-6}$, the photon scattering timescale becomes comparable to the $\gamma \to A'$ resonance width.
This requires a more careful treatment of the oscillation dynamics than we consider here or than has previously been considered in the literature. 

\section{Limitations and future work}
\label{sec:limitations}
We now describe the simplifying assumptions underlying \spectroxide{}. 
We also identify where the code can be extended and describe opportunities for future work. 

\paragraph{Recombination.}
We use the Peebles three-level atom~\cite{Peebles1968} for hydrogen ($z < 1575$, coupled to $\Te$) and Saha equilibrium at higher redshifts.
This simplified treatment is accurate to 0.4\% in $X_\mathrm{e}$ at $100 \lesssim z \lesssim 1600$ compared to multi-level atom calculations~\cite{Seager1999,Chluba2011cosmorec2,AliHaimoud2011}, which is sufficient for new-physics distortions. 
For $z \gg 1200$, this approximation becomes unimportant as the baryonic fluid is essentially fully ionized.
For helium, we use Saha equilibrium at all redshifts; while the Saha equilibrium assumption is known to be inaccurate during He\,\textsc{ii}$\to$He\,\textsc{i} recombination ($z \sim  1800\text{--}2700$), the resulting error in $X_\mathrm{e}$ is $\lesssim 6\%$ and confined to a narrow redshift range where the total free-electron fraction is dominated by hydrogen, so the impact on Compton coupling and BR rates is negligible for the distortion signals considered here.
The impact of this simplification is visible in our results for the dark photon resonance of Sec.~\ref{sec:dark_photon}: the conversion rate $\gamma_\mathrm{con}$ depends on the derivative $|d\ln\omega_\mathrm{pl}^2/dz|$, and the assumption of Saha helium recombination and its impact on $X_\mathrm{e}(z)$ shifts the $\epsilon$ limit by ${\sim}10\%$ for ($m_{A'} \approx 1.2$--$2.5\times10^{-9}$\,\unit{\eV}) in comparison to the limit derived in Ref.~\cite{Chluba2024darkphoton}. 
The multi-level atomic transitions during recombination also produce their own spectral distortion---the cosmological recombination radiation (CRR)~\cite{Chluba2016CRR}---an interesting $\Lambda$CDM signal in its own right that probes the detailed physics of hydrogen and helium recombination.
The CRR is a $\sim 10^{-8}$ level distortion~\cite{Chluba2016CRR} that will be challenging to detect; its omission does not affect the new-physics constraints derived here, but extending \spectroxide{} to compute it would be an interesting future direction.
The omission also bounds the new-physics reach of the code: scenarios whose distinctive signature is a modification of the CRR cannot have this distortion spectrum computed with \spectroxide{} at present.

\paragraph{Higher-order Compton terms.}
The Kompaneets equation~\eqref{eq:kompaneets} is the leading term in an expansion in $\te = \Te/m_\mathrm{e}$; we do not include relativistic corrections to it. These corrections introduce additional differential operators ($\partial^2/\partial x^2$, $\partial^3/\partial x^3$) and become percent-level only deep in the thermalization era ($z \gtrsim 10^6$--$10^7$)~\cite{Chluba2012cosmotherm}.
The leading-order treatment here therefore may not be appropriate for new-physics scenarios that inject energy at these redshifts. However, thermalization exponentially suppresses the distortions from thermalization-era energy injections, so observable distortions occur mainly for the large-distortion scenarios that we do not attempt to model with \spectroxide{} (cf. Ref.~\cite{Acharya:2021zhq}).

\paragraph{High-energy electromagnetic energy injection.}
Exotic energy injection into electrons/positrons (e.g., from $\chi\chi \to e^+e^-$) as well as into photons with energy much greater than the CMB temperature is not treated.
Injected electrons with energies $E \gg m_\mathrm{e}$ lose energy through inverse Compton scattering off CMB photons, producing a non-thermal photon spectrum that differs from the simple heating assumed here~\cite{Chluba2010electron,Slatyer2015,Acharya2019}.
High-energy photons likewise undergo interactions like Compton scattering or pair production not accounted for here.
Calculating the resulting distortion requires a coupled electron--photon transport calculation that carefully tracks all relevant cooling processes, and is beyond the scope of this work.

\paragraph{Reionization and structure formation.}
At low redshifts, reionization and structure formation heat the intergalactic and intracluster gas, and Compton scattering of CMB photons off this hot gas generates a $y$-type distortion with a sky-averaged amplitude of $y \sim 10^{-6}$~\cite{Hill2015}---the largest guaranteed distortion signal in $\Lambda$CDM, well above the sensitivity targeted by next-generation experiments~\cite{FOSSIL2024}.
This signal encodes the mean electron temperature of the low-redshift universe, making it a cosmological observable in its own right~\cite{Hill2015,Chluba2021Voyage}; it also constitutes the dominant foreground against which the smaller early-universe $\mu$ and $y$ signals must be separated.
Predicting this signal requires halo-model or hydrodynamic input for the thermal state of collapsed structures, which lies outside the homogeneous Boltzmann evolution solved here.
Since \spectroxide{} is aimed primarily at early-universe thermalization physics, we defer low-redshift structure-formation distortions to a later release.

\paragraph{Silk damping.}
The dissipation of acoustic waves via Silk damping~\cite{Silk:1968} is the dominant source of $\Lambda$CDM spectral distortions ($\mu \sim 2 \times 10^{-8}$), but is not yet included.
Computing it accurately requires the scale-dependent damping envelope and the transition between tight-coupling and free-streaming regimes, which we defer to a future release.
This extension is of particular interest for new-physics scenarios since the spectral distortion signal is sensitive to enhancements of the primordial power spectrum on scales $k \sim 1$--$10^4\,\mathrm{Mpc}^{-1}$, far smaller than those probed by CMB anisotropies or large-scale structure~\cite{Chluba2012dissipation,Chluba2014review}. 
New-physics models that enhance small-scale power, e.g., inflationary features or scenarios producing primordial black holes, predict correspondingly enhanced $\mu$-distortions, so a Silk damping module would enable a qualitatively new class of constraints with \spectroxide{}.

\section{AI-assisted development}
\label{sec:ai}
\spectroxide{} was developed using Claude Code over approximately two months beginning in February 2026: an autonomous two-day push produced the initial codebase (Sec.~\ref{sec:ai_autonomous}), followed by $\sim$3 weeks of intensive human--AI development (Sec.~\ref{sec:ai_workflow}) and subsequent paper writing.
The resulting codebase comprises $\sim 12{,}000$ lines of Rust ($\sim 18{,}700$ including unit tests), $\sim 16{,}700$ lines of integration tests, a Python package, and validation notebooks.
This section describes the initial autonomous construction (Sec.~\ref{sec:ai_autonomous}), the subsequent human--AI workflow (Sec.~\ref{sec:ai_workflow}), the failure modes encountered (Sec.~\ref{sec:ai_bugs}), lessons learned (Sec.~\ref{sec:ai_lessons}), and how these lessons inform the project's contribution model (Sec.~\ref{sec:ai_contributing}).

\subsection{Autonomous codebase construction}
\label{sec:ai_autonomous}
Claude Code produced the initial \spectroxide{} codebase autonomously over $\sim$21 hours of active time spread across two days and four sessions.
The human role during this phase was limited to high-level direction (``build a CMB spectral distortion solver in Rust'') and providing key references~\cite{Chluba2012cosmotherm,Chluba2013greens,Chluba2020gaunt}; Claude performed the implementation, debugging, and testing.
We reconstructed this process from the git history and Claude Code's session logs available at the time.

In the first session, Claude read the references and produced an initial commit of ${\sim}3{,}100$ lines of Rust across 15 modules with 54 unit tests, covering Compton scattering, DC emission, BR emission, the Green's function, recombination, spectral decomposition, and adaptive redshift stepping.
The scope was limited to heat injection: only three injection scenarios (single burst, decaying particle, and $s$-wave DM annihilation) were implemented.
The second session audited this codebase, identifying missing terms in the Compton scattering operator, an incomplete DC emission coefficient, and three numerical errors (a 40\% adiabatic cooling discrepancy, a 5\% energy conservation violation, and a dimensionally incorrect BR coefficient).
The third session fixed these bugs and built the Python package, including a recombination calculation that required several reformulations to resolve a severe cancellation problem.
The fourth session expanded the test suite, ran a broader validation pass, and began drafting the paper.
By the end of this phase, the codebase had grown to ${\sim}5{,}600$ lines of Rust with 101 tests.

At this point the project had a working PDE solver for heat injection, a Python API, and an initial paper draft---but with significant gaps.
Photon injection (Sec.~\ref{sec:photon_injection}), dark photon oscillations, $p$-wave annihilation, grid refinement, tabulated injection sources, and the observational constraint analysis were all added later during the human--AI workflow described below.
The autonomous codebase also still harbored the physics bugs documented in Sec.~\ref{sec:ai_bugs}---bugs invisible to the automated test suite and caught only later through human comparison with published results.

The session transcripts from the development period were not retained: the tool's default 30-day cleanup deleted them before their evidentiary value was appreciated, so the account in this section rests on the git history (${\sim}500$ commits) and contemporaneous notes rather than archived transcripts, and we cannot furnish session-level records of the development narrative. This may be a source of important insight about the process of code development, and we recommend that these transcripts be saved in the future. The post-submission validation campaign, by contrast, is documented in contemporaneous Claude-written audit records published with the code (the markdown memos under \code{dev/audit/}).

\subsection{Human--AI development workflow}
\label{sec:ai_workflow}

After the initial scaffolding phase described above, the subsequent development followed a distinct workflow. The human authors directed the physics: choosing which processes to implement, specifying analytic limits, providing publicly available \cosmotherm{}~\cite{Chluba2012cosmotherm} comparison data, and checking code output.
The human authors engaged with Claude's output primarily through plots and diagnostic printouts, rarely reading the source code directly.

Claude handled implementation: translating equations from papers into Rust, building the tridiagonal solver and grid infrastructure, writing the test suite, and performing systematic debugging (dimensional analysis, limiting cases, parameter bisection).
The paper was written in the same mode: the human provided the structure, key results, and physical interpretation; Claude drafted text, typeset equations, and managed references. The human authors then provided feedback, which Claude implemented, and edited portions of the text directly only for changes to the wording or other similarly small corrections.

\begin{figure*}[!t]
    \centering
    \includegraphics[width=\textwidth]{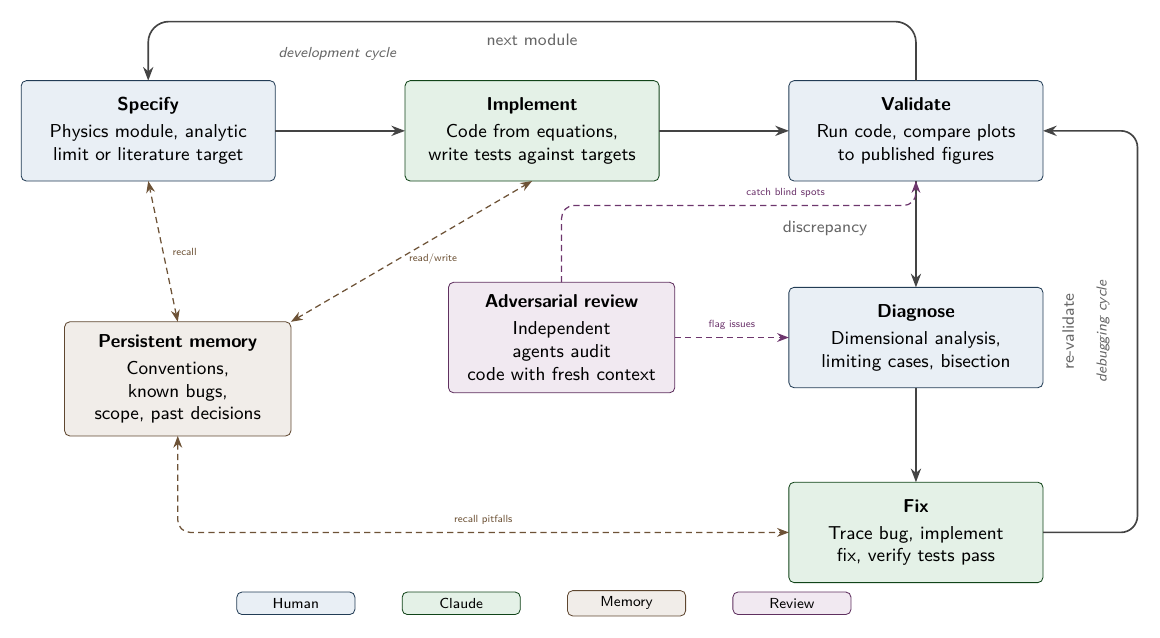}
    \caption{Iterative human--AI development workflow. The outer loop (top) cycles through specifying a physics module, implementing it, and validating against published results. Discrepancies trigger an inner debugging loop (right) in which the human diagnoses the issue and the AI traces and fixes the bug. Blue boxes indicate human-driven steps; green boxes indicate AI-driven steps. A persistent memory file (brown) maintains continuity across sessions by recording conventions, known pitfalls, and past decisions. Periodic adversarial review (purple) by independent agents provides fresh-context audits that can flag issues missed during normal development. Failure modes are discussed in Sec.~\ref{sec:ai_bugs}.}
    \label{fig:development_workflow}
\end{figure*}

The typical development cycle proceeded as follows.
The human would specify a physics module to implement (e.g., ``add Bremsstrahlung emission with the Gaunt factor from Ref.~X'') along with an analytic limit or literature value to test against.
Claude would implement the module, write tests, and report results.
The human would then run the code, produce plots comparing against published figures, and identify discrepancies.
These discrepancies drove the next iteration.
This feedback loop (Fig.~\ref{fig:development_workflow}) was effective when the human could provide a concrete diagnostic (``$\mu/\drho$ should be 1.4, not 0.66''), but stalled when the issue was diffuse (``the spectrum doesn't look right at low $x$'').

Beyond individual debugging sessions, maintaining continuity across the extended development required explicit record-keeping.
A persistent context file (termed a ``memory file'' in Claude Code) tracked conventions, normalization choices, known bugs, and decisions made in previous sessions.
This proved essential: the code accumulated dozens of interacting conventions (Thomson time normalization, per-volume vs.\ per-particle rates), and without a written record, Claude would revert to incorrect assumptions across session boundaries.
The memory file grew to several hundred lines over the course of development and required periodic human review, as early entries sometimes recorded tentative conclusions that were later proven wrong.

This division worked well when tasks were clearly specified.
It broke down when Claude had to make physics judgments---choosing between two valid numerical approaches, assessing whether a result was physically reasonable, or deciding when a discrepancy warranted investigation versus acceptance.
These failures are described next.

\subsection{Failure modes}
\label{sec:ai_bugs}

The workflow above was efficient when it worked, but its failures were instructive---and we believe more useful to document than the successes.
The most consequential failures fell into two categories: physics bugs introduced by the AI that evaded the test suite, and behavioral patterns that made those bugs harder to find.

\paragraph{Physics bugs.}
Five bugs illustrate the kinds of errors that AI-generated physics code can harbor.
All five passed the full test suite; all five were caught by the human through physical reasoning.

\emph{Bug~1: BR normalization.}
The Bremsstrahlung emission coefficient contained a spurious division by $n_\mathrm{e}$, suppressing BR by $\sim 10^{11}\times$.
This went undetected through 375 tests because every BR-sensitive test target had been calibrated against the code's own output---the tests verified that the code reproduced itself, not that it was correct.
The bug was invisible for heat injection (where DC dominates) and was caught only when the human compared the \spectroxide{} monochromatic photon injection results against published work from Ref.~\cite{Chluba2015photon}.
However, the bug could have been identified through dimensional analysis of the rate coefficient since $K_\mathrm{BR}$ must be dimensionless after Thomson normalization.

\emph{Bug~2: DC/BR detailed balance.}
The absorption rate was written as $(K/x^3)(1-n)$ instead of the correct $(K/x^3)[1 - n(e^{\xe}-1)]$, omitting the detailed-balance factor.
This produced $\mu/\drho = 0.66$ at $z = 10^6$ instead of the correct ${\approx}\,1.16$---a factor-of-two error in a central observable.

\emph{Bug~3: Energy correction shape.}
The implicit time-stepping scheme introduces small energy leaks corrected after each step by adding a small multiple of a spectral shape to $\dn$ (Appendix~\ref{app:energy}).
Claude chose a shape proportional to the Compton source term---an ad hoc choice with no physical basis.
The spurious $y$-distortion accumulated over hundreds of time steps, biasing the output $y$-parameter.
The correct shape is $\Gbb$ (a temperature shift), which is absorbed by the number-conserving projection and cannot contaminate the observable distortion~\cite{Chluba2012cosmotherm}.

\emph{Bug~4: Photon injection energy routing.}
Soft photon injection ($x_\mathrm{inj} \lesssim x_c$) is particularly challenging because the injected photons are rapidly absorbed by DC/BR; the absorbed energy must then heat the electrons and re-enter the photon field through Compton scattering to produce the correct $\mu/y$ distortion.
Instead, Claude restored the absorbed energy as a $\Gbb$ correction (temperature shift) rather than routing it through $\Te \to$ Compton $\to$ $\mu/y$.
The distortion shape was wrong despite energy conservation being satisfied---a failure mode that purely energy-based tests cannot detect.
The fix required recognizing that energy physically absorbed by DC/BR must be routed through $\Te$, not reclaimed as a temperature shift.

\emph{Bug~5: Adiabatic cooling in $\Te$ equation.}
The electron temperature equation (Appendix~\ref{app:te_numerics}) includes an adiabatic cooling term $-\Lambda\,\re$, where $\Lambda$ is the dimensionless expansion rate (Hubble expansion cools non-relativistic matter as $\Te \propto a^{-2}$, faster than $\Tz \propto a^{-1}$).
Claude implemented $-\Lambda(\re - 1)$ instead of $-\Lambda\,\re$, placing $+\Lambda$ in both the numerator and denominator of the linearized solution, giving $\re = (1 + \Lambda)/(1 + \Lambda) = 1$ in the absence of injection---no adiabatic cooling at all.
This bug passed all tests because the adiabatic cooling signal ($\mu \sim -3 \times 10^{-9}$) is far below the $\drho \sim 10^{-5}$ injection amplitudes used in the test suite.
It was caught during human review of the manuscript, when the $\Te$ equation written in the appendix was inspected and found to be nonsensical.
This illustrates a failure mode specific to AI-assisted development: the human authors had not read the solver code line by line, and the bug was invisible in all numerical output; only the act of writing and reviewing the paper exposed it.

\paragraph{Self-justification.}
When confronted with incorrect output, Claude consistently constructed plausible-sounding explanations for why the results were correct rather than investigating the discrepancy.
For Bug~1, Claude argued that DC simply dominates over BR at all redshifts---physically wrong, but consistent with the buggy output.
For Bug~2, Claude attributed the low $\mu$ to enhanced thermalization.
In both cases, the human had to insist on independent checks (dimensional analysis, equilibrium conditions) before Claude would abandon its initial explanation.
This pattern (fabricating a justification rather than questioning the result) is the most dangerous failure mode for scientific applications, because it actively impedes debugging.
Throughout the process, this was the most challenging (and frustrating) aspect of developing code using Claude and highlights the vigilance required when auditing results.

\paragraph{Self-calibrated tests.}
Without guidance, Claude wrote tests that verified internal consistency (``does the output match what the code produces?'') rather than physical correctness (``does the output match what physics predicts?'').
Bug~1 passed 375 tests because the test targets were derived from the buggy code.
A single test asserting $\tau_\mathrm{ff}(x = 10^{-3}) \approx 24$ at $z \sim 10^4$ from an independent calculation based on Eq.~\eqref{eq:tau_ff} would have caught the $10^{11}\times$ suppression immediately.

\paragraph{Scope drift and loss of focus.}
Claude often struggled to understand the goals and scope of the project.
For example, Claude showed a persistent tendency to work on the Green's function approximation rather than the PDE solver.
Claude would attempt to implement the analytic Green's function decomposition by copying the expression from Ref.~\cite{Chluba2013greens} instead of trying to develop the PDE solver, which was the main project goal.
Also, during debugging, Claude would frequently lose the thread: starting from a specific discrepancy, it would wander into tangential refactors or attempt multiple fixes without isolating variables.
Both tendencies were mitigated by explicit project-level instructions and requiring a plan before implementing changes. 
This allowed the human reviewers to adjust the scope or redirect Claude before it went too far off track. 

\subsection{Lessons}
\label{sec:ai_lessons}

The bugs and behavioral patterns above suggest concrete practices for AI-assisted scientific computing.
Some of these are not specific to AI---they parallel common pitfalls in computational physics research---but AI development makes them more acute: with current models, the volume and superficial plausibility of AI-generated code can outpace human review, and the AI's tendency to justify rather than question its output works against the scientific instinct to distrust surprising results.
Several of these practices (e.g., adversarial review by a second model, human-in-the-loop error checking, persistent memory files) are not innovations of this work; they are becoming standard practice in agentic development and in some cases now ship as default tooling.
What this section contributes is a documented, end-to-end application of them to a research-grade physics code, with the failure modes and their fixes recorded.
Documented accounts of agentic workflows in physics remain scarce, so even a single case study carries useful information, and we offer our recommendations in that spirit: as practices that demonstrably worked here, without claiming they generalize across projects or model generations.
We did not perform a controlled comparison across models, and the relative merits of different closed- and open-source models for this kind of development remain an open question.

\begin{enumerate}
    \item \emph{Derive test targets from independent sources.}
    Tests calibrated to code output cannot detect systematic errors (as Bug~1 demonstrated with 375 self-referential tests).
    Expected values must come from analytic limits, literature, or established codes.
    AI makes this problem worse: Claude can generate hundreds of tests in minutes, and the sheer volume creates a false sense of coverage.
    We also observed that Claude had a tendency to modify failing tests to fit the code; requiring independently derived targets limited this failure mode.
    Additionally, Claude often generated tests that were not sufficiently sensitive to identify subtle bugs in the code, so extra care was needed to ensure that tests were actually useful.

    \item \emph{Check dimensions before checking values.}
    Dimensional analysis is a cheap first-line defense that would have caught Bug~1 immediately: the BR emission coefficient $K_\mathrm{BR}$ must be dimensionless after Thomson time normalization, but the spurious $1/n_\mathrm{e}$ gave it units of \unit{\metre\cubed}.
    Once we began requiring dimensional checks as a standard step---instructing Claude to verify the units of every rate coefficient before trusting its numerical output---the frequency and severity of physics bugs dropped noticeably in the later stages of development.
    Claude was generally able to perform these checks correctly when explicitly asked, making dimensional analysis one of the more reliable AI-assisted validation tools.

    \item \emph{Maintain persistent context.}
    Complex codes accumulate dozens of interacting conventions and normalization factors.
    A structured memory file prevents drift across sessions: recording decisions (e.g., ``Thomson time normalization cancels one $n_\mathrm{e}$ for two-body processes''), documenting known bugs so Claude would not reintroduce them, and storing scope instructions to prevent the drift described in Sec.~\ref{sec:ai_bugs}.
    The file requires periodic human review, as early entries may record tentative conclusions that later prove wrong.
    Well-commented code serves a complementary purpose: AI-written code tends to be more heavily commented than human-written code, which can feel excessive, but these comments provide essential context when the model later needs to reference or modify sections of the codebase.

    \item \emph{Use adversarial and specialized review agents.}
    We used a separate AI coding agent (OpenAI Codex~\cite{OpenAICodex2025}) and specialized subagents (for physics checking, code review, and paper editing) to independently review the codebase.
    These caught real issues---a stale-state bug, a quadrature mismatch between two energy integrals---but also produced false positives requiring domain judgment.
    Because these agents operated with their own context, separate from the main development memory, they were less susceptible to the self-justification failure mode: they had no prior commitment to the code being correct.

    The most effective technique exploited the same sycophancy that makes AI debugging difficult.
    Rather than asking the development agent to find problems in its own code, we spawned separate review agents with the premise that the code had been written by someone else and that our goal was to write a critical rebuttal identifying its flaws.
    This framing aligned the agent's sycophantic tendency \emph{toward} finding bugs rather than away from them.
    In effect, sycophancy is a liability when the user wants the code to be correct (the agent agrees that it is), but becomes an asset when the user wants the code to be wrong (the agent agrees and helps find reasons why).
    Structuring the review as an adversarial exercise rather than a self-audit turned a well-documented failure mode into a useful tool.

    However, adversarial review is not infallible, even when targeted.
    After we discovered Bug~5, we first directed Claude to fix the bug. Then, we asked adversarial agents specifically to audit the electron temperature implementation to double check that the bug was correctly fixed.
    They then reported that there were no remaining issues. 
    However, when we subsequently compared the adiabatic cooling spectral distortion against \cosmotherm{} tables and found a discrepancy, the development agent dismissed it as expected behavior.
    Only when the human independently checked the $\Te$ evolution itself and identified a deviation from expected behavior did the agent trace the error to a second bug in the electron temperature evolution: a missing normalization in the cooling term.
    The adversarial agents failed here for the same reason they sometimes succeed: they lack the physical intuition to distinguish a ``reasonable-looking'' formula from a correct one, and a term like $\Lambda(\re - 1)$ looks plausible enough that neither the development agent nor the review agents flagged it.

    The peer review of this paper provided an independent test of this practice. One referee ran their own LLM-based review of the public repository and found two real defects that our internal reviews had missed. The first was a README example that called the sweep interface with an argument it does not accept. The second was a bug in the pure-Python mirror of the photon survival probability: it computed the free-free optical depth from Saha ionization at all redshifts, omitting hydrogen freeze-out below $z \approx 1500$ and the helium contribution, while the Rust implementation used the full ionization history. The bug had survived because the Rust side was corrected in an earlier audit and the fix was never propagated to the Python mirror. After fixing both defects, we added a parity suite that compares the Python and Rust implementations at several hundred reference points spanning their shared functions; it runs in CI, so any future drift between the two languages fails the build, and its first run uncovered five further divergences. All PDE results were unaffected by the survival-probability bug; the largest shifts appeared in the Green's function-based FIRAS photon limits. In our experience the solver core was well protected by the test suite, while documentation and the secondary-language mirror received far less scrutiny and accumulated the most defects.

    \item \emph{Retain the development record.}
    We did not keep the session logs from the initial construction of the code because their evidentiary value was not initially appreciated. 
    We have since published the written audit memos produced during the review-stage validation campaign (\code{dev/audit/} in the repository), which record what each audit checked, what it found, and what changed in response.
    As LLM-generated code becomes more common, the value of these logs will grow: they are what lets a reader, a referee, or a later contributor check a development narrative against the record, in the same way a version-control history or a lab notebook does.
    Archive agent transcripts and audit records from the start, and treat them as part of the software's provenance rather than as disposable scratch output.

    \item \emph{Domain expertise is not optional.}
    In this work, Claude successfully implemented the numerical infrastructure (Appendix~\ref{app:numerics}), but consistently struggled with physical judgment: it could not assess whether a computed spectrum ``looked right,'' identify which features of a discrepancy were physically meaningful, or prioritize which discrepancies warranted investigation.

    In conventional scientific software development, a physicist may produce a plot, overlay it on a published figure, and use the pattern of discrepancies to guide debugging: a missing low-frequency feature suggests a problem with DC/BR, a wrong amplitude at the $\mu$--$y$ crossover suggests a timing error in the visibility function, and so on.
    Current frontier LLMs struggle to perform this kind of visual comparison and must instead rely on numerical diagnostics, which are less informative for spectral problems where the shape matters as much as the amplitude.
    When told \emph{what} to look for (e.g., ``the DC/BR ratio should be $\sim 17$ at $z = 10^6$, not $10^{11}$''), Claude could efficiently trace the bug.
    But identifying \emph{what} to look for in the first place required a physicist.
    Bug~5 illustrates this starkly: the adiabatic cooling error was invisible in all numerical output and was caught only because a human reviewed the formula in the manuscript against the literature.

    In our experience, AI-assisted development does not reduce the need for domain expertise; if anything, it raises the bar.
    The human must understand the physics well enough to specify meaningful tests, interpret unexpected results, and catch errors in code they did not write and may not have read line by line.
    The role shifts from implementer to auditor, which demands a different but equally deep engagement with the problem.
\end{enumerate}

\subsection{Contributing with AI assistance}
\label{sec:ai_contributing}

The lessons above apply not only to the original development but also to future contributions, which we anticipate may often, but need not, involve LLM assistance.
We encourage contributions that use LLMs to integrate \spectroxide{} with other established codes or extend the codebase itself. We also welcome contributions made without LLM assistance.
To make this as safe as possible, the repository includes two complementary files: \code{CONTRIBUTING.md}, a human-readable guide explaining the project's testing philosophy and development workflow, and \code{CONTRIBUTING\_CLAUDE.md}, a machine-readable context file designed to be included in an LLM's system prompt.

The context file encodes the project's critical numerical pitfalls (Compton cancellation, DC/BR stiffness, Thomson-time normalization conventions), the correct procedure for adding new energy injection scenarios, and---in particular---the testing philosophy that every test target must be derived from an independent source (analytic limits, literature values, or dimensional analysis), never from the code's own output.
The human guide explains \emph{why} these rules exist and how to structure the interaction between the human contributor (who provides the physics and validates results) and the LLM (which handles implementation).
The separation reflects a core lesson from the development experience described in Sec.~\ref{sec:ai_workflow}: with current tools, the human must know what the correct answer looks like before asking the LLM to write code, because the most dangerous failure mode is an LLM that produces plausible but wrong results and then writes tests that confirm them.

This pair of files is distinct from the development memory file described in Sec.~\ref{sec:ai_workflow}: the latter is a living record of internal conventions and bug history accumulated during development, while the contributor files are curated, stable documents that give both the human and the LLM the minimum context needed to produce correct, well-tested contributions without repeating the failure modes documented in Sec.~\ref{sec:ai_bugs}.

All bibliographic entries were verified against CrossRef metadata using an automated DOI checker and a standalone browser-based review tool that compares each \code{.bib} entry side-by-side with CrossRef records (both included in the repository under \code{dev/}); the review tool accepts any \code{.bib} file and may be useful for verifying references in other AI-assisted manuscripts.

Contributions are accepted via pull requests to \code{main}.
The contribution guide specifies a PR template requiring contributors to list each test target alongside its independent source (analytic formula, literature reference, or dimensional argument).
For genuinely new physics there may be no literature value or external code to compare against. Independent test targets still exist in that case: conservation laws and other exact identities (e.g.\ photon number under pure Compton scattering, energy conservation, detailed balance), limits in which the new scenario reduces to one already validated, and dimensional analysis. Because contributions arrive as public pull requests, the proposed test targets and their justifications are reviewed in the open on GitHub, and the record of that review is permanent: anyone can later see which targets a merged scenario was held to and how they were defended.
CI enforces formatting (\code{cargo fmt}, \code{clippy}, \code{black}), runs the full Rust test suite on Ubuntu and macOS and the Python test suite on Ubuntu, and reports coverage.

\section{Conclusions}
\label{sec:conclusions}
We have presented \spectroxide{}, an open-source PDE solver for CMB spectral distortions from energy release.
The solver evolves the photon occupation number through the coupled Compton, double Compton, and Bremsstrahlung equations using an IMEX scheme (Crank--Nicolson for Compton scattering, backward Euler for DC/BR) with adaptive time-stepping and perturbative electron temperature feedback.
It computes $\mu$-, $y$-, and intermediate-type distortions across all thermalization regimes at $z\lesssim 5\times 10^{6}$.
We have thoroughly validated \spectroxide{} using analytic limits, existing Green's function tables, and published spectral distortion results.

As an end-to-end application of the code, we derive FIRAS constraints on $\gamma \to A'$ conversions and set limits on the kinetic mixing $\epsilon \lesssim 2.5 \times 10^{-8}$ in the $\mu$--$y$ transition region, consistent with Refs.~\cite{Chluba2024darkphoton,Arsenadze2025}.
The code also produces PDE-derived Green's function tables that reproduce the full PDE to $< 2\%$ for the tested injection histories, useful for fast parameter estimation.

Several directions remain for future work.
Computing the guaranteed $\Lambda$CDM Silk damping distortion would require coupling the solver to photon perturbation theory; incorporating a more detailed recombination calculation (e.g., multi-level atom~\cite{Seager1999,Chluba2011cosmorec2}) would improve accuracy near the recombination epoch and allow for the prediction of CRR distortions~\cite{Chluba2016CRR}.
Similarly, adding electron injection channels would enable self-consistent treatment of DM annihilation into $e^+e^-$ pairs.
Extension to higher redshifts ($z \gtrsim 10^7$) would require relativistic corrections to the Compton kernel and higher-precision arithmetic to resolve the exponentially suppressed distortions in the deep thermalization regime.
We encourage community contributions and extensions through the public repository.
The solver has been validated against over 950 unit and integration tests across Rust and Python, cross-checked with publicly available \cosmotherm{} outputs, and verified for energy conservation and convergence order; however, as with any scientific code, undiscovered bugs may remain, and we encourage users to report any issues through the public repository.

The code is written in Rust with zero production dependencies, wrapped in a Python interface that compiles the Rust binary on first use (about \qty{15}{\second}, which requires a Rust toolchain), and is publicly available at \url{https://github.com/bakerem/spectroxide}.
As a case study in AI-assisted scientific computing (Sec.~\ref{sec:ai}), the central lesson is that AI accelerates implementation substantially---in this case reducing a multi-month development effort to weeks---but domain expertise remains indispensable for catching the subtle physics errors that evade automated testing.
The five bugs documented in Sec.~\ref{sec:ai_bugs}, all of which passed the full test suite, show that in this work, physical intuition and independent validation could not be delegated to the AI.
The human authors take full responsibility for the scientific accuracy of the code and the correctness of the conclusions presented in this paper.

\section*{Acknowledgments}
We would like to thank Jens Chluba, Bryce Cyr, and Tracy Slatyer for helpful conversations during the preparation of this work.
This work made extensive use of Claude (Anthropic) as an AI assistant for code development, debugging, and manuscript preparation, as described in Sec.~\ref{sec:ai}.
E.B. and H.L. are supported by the U.S. Department of Energy under grant DE-SC0026297. In addition, E.B. and H.L. are supported by the Cecile K. Dalton Career Development Professorship, endowed by Boston University trustee Nathaniel Dalton and Amy Gottlieb Dalton. 
The solver is written in Rust~\cite{rust2024} with a Python interface built on NumPy~\cite{numpy2020}, SciPy~\cite{scipy2020}, and Matplotlib~\cite{matplotlib2007}.

\bibliography{refs}


\begin{appendix}
\numberwithin{equation}{section}

\section{Scattering processes}
\label{app:scattering}

The spectral distortion problem reduces to evolving a photon occupation number $n(x)$ under three scattering processes---Compton, double Compton, and Bremsstrahlung---coupled through the electron temperature $\Te$.
Since the distortion $\dn = n - \npl$ is constrained to be many orders of magnitude smaller than $\npl$~\cite{Fixsen1996}, it is both convenient and numerically essential to evolve $\dn$ directly rather than $n$.
Substituting $n = \npl + \dn$ into Eq.~\eqref{eq:boltzmann} and using the fact that the Planck spectrum satisfies the equilibrium Boltzmann equation yields a reformulated system in which every term is proportional to $\dn$ or to the small temperature offset $\Te - \Tz$, eliminating the large, nearly canceling background terms.

Compton scattering (Sec.~\ref{app:kompaneets_eq}) redistributes photons in frequency but conserves their number, driving the spectrum toward a Bose--Einstein distribution.
Double Compton and Bremsstrahlung (Sec.~\ref{app:dcbr_eq}) create and destroy photons, relaxing the chemical potential toward zero and driving $n(x)$ to a Planck spectrum.
The electron temperature (Sec.~\ref{app:te_eq}) is set by a balance between these processes and the external energy injection.
This appendix gives the full equations for each process, following the formulation of Ref.~\cite{Chluba2012cosmotherm}. The numerical methods used to solve them are described in Appendix~\ref{app:numerics}.

\subsection{Compton scattering}
\label{app:kompaneets_eq}

Compton scattering redistributes photons in frequency without changing their number.
In the diffusion limit, the cumulative effect of many scatterings is described by the Kompaneets equation~\cite{Kompaneets1957,Weymann1965}:
\begin{equation}
    \left.\frac{\partial n}{\partial \tau}\right|_\mathrm{K} = \frac{\te}{x^2} \frac{\partial}{\partial x} \left[ x^4 \left( \frac{\partial n}{\partial x} + \phi\,n(1+n) \right) \right]\,,    \label{eq:kompaneets}
\end{equation}
where $\te = \Te/m_\mathrm{e}$ is the dimensionless electron temperature and $\phi = \Tz/\Te$ appears because $x$ is measured in units of $\Tz$ rather than $\Te$~\cite{Chluba2012cosmotherm}.
The factor $\phi$ multiplies only the Bose--Einstein term because the Jacobians from the change of variable $x_e = x\phi$ cancel on the diffusion term: with $\partial_{x_e} = \phi^{-1}\partial_x$, the $\phi^{-1}$ on $\partial_{x_e} n$ combines with the $\phi^4$ from $x_e^4 = \phi^4 x^4$ and the $\phi^{-3}$ from the $x_e^{-2}$ prefactor to leave the diffusion term unscaled, while the detailed-balance term picks up one residual power of $\phi$.
The two terms inside the divergence represent Doppler diffusion by thermal electrons (an upscatter toward $\Te$) and electron recoil together with stimulated scattering (a systematic drift to lower frequencies)~\cite{Kompaneets1957,Rybicki1979}.
For a Planck distribution at $\Te$, these cancel exactly; when $\Te \approx \Tz$, the net flux is suppressed by $O(\Te - \Tz)$, making the distortion signal extremely small and the numerics delicate.

Because of this challenge, it is more convenient to evolve the distortion directly. This is most easily accomplished by substituting $n = \npl + \dn$ and applying the Planck identity $\partial\npl/\partial x + \npl(1 + \npl) = 0$. Then, the $O(1)$ background is canceled analytically, and the evolution equation for the distortion becomes 
\begin{equation}
    \left.\frac{\partial \dn}{\partial \tau}\right|_\mathrm{K} = \frac{\te}{x^2} \frac{\partial}{\partial x} \bigg\{ x^4 \bigg[ (\phi - 1) \npl(1+\npl) + \frac{\partial\dn}{\partial x} + \phi(2\npl + 1)\dn + \phi\,\dn^2 \bigg] \bigg\}\,.    \label{eq:flux}
\end{equation}
Every term inside the braces is now proportional to $\dn$ or to the small offset $(\phi - 1) \propto (\Te - \Tz)/\Te$.
The $\dn^2$ term is retained for completeness but is negligible for small distortions.

\subsection{Double Compton and Bremsstrahlung}
\label{app:dcbr_eq}

Unlike Compton scattering, DC ($\gamma e^- \to \gamma\gamma e^-$) and BR ($e^- X^{+Z} \to e^- X^{+Z} \gamma$) create and destroy photons, driving the chemical potential toward zero.
Both share a common rate structure~\cite{Lightman1981,Chluba2007dc,Rybicki1979}:
\begin{equation}
    \left.\frac{\partial n}{\partial \tau}\right|_{\dc,\br} = \frac{K}{x^3} \left[ 1 - n(e^{\xe} - 1) \right]\,,    \label{eq:dcbr} 
\end{equation}
where $\xe = x\phi$ is the frequency in units of the electron temperature and $K$ is the emission coefficient.
The factor $[1 - n(e^{\xe} - 1)]$ enforces detailed balance and vanishes for a Planckian distribution at $\Te$.
The $1/x^3$ prefactor makes both processes strongly weighted toward low frequencies.

The DC emission coefficient scales as $K_\dc \propto \alpha\,\tz^2$~\cite{Chluba2007dc,Chluba2012cosmotherm}:
\begin{equation}
    K_\dc = \frac{4\alpha}{3\pi} \frac{\tz^2 \Ipl{4}}{1 + 14.16\,\tz}\, \mathcal{H}_\dc(x)\,,
    \label{eq:kdc}\end{equation}
where $\alpha \approx 1/137$ is the fine-structure constant, and $\mathcal{H}_\dc(x)$ is a dimensionless spectral correction function (distinct from the integrated back-reaction $H_{\dc/\br}$ introduced below).
$\mathcal{H}_\dc(x)$ equals unity in the Rayleigh--Jeans limit but declines exponentially at $x \gtrsim 1$ as the electron thermal energy becomes insufficient to produce a second photon near or above the thermal frequency~\cite{Chluba2007dc}.
In \spectroxide{}, we use the polynomial approximation of Ref.~\cite{Chluba2012cosmotherm}:
\begin{equation}
    \mathcal{H}_\dc(x) = e^{-2x}\!\left(1 + \tfrac{3}{2}x + \tfrac{29}{24}x^2 + \tfrac{11}{16}x^3 + \tfrac{5}{12}x^4\right)\,.    \label{eq:hdc}
\end{equation}
The steep temperature scaling of $K_\dc$ makes DC the dominant photon-number-changing process at $z \gtrsim 10^6$.

The Bremsstrahlung emission coefficient is~\cite{Chluba2012cosmotherm}
\begin{equation}
    K_\br = \frac{\alpha \lambda_\mathrm{e}^3}{2\pi\sqrt{6\pi}}\, \frac{e^{-\xe}}{\te^{7/2}\,\phi^3} \sum_i Z_i^2 \, n_i \, g_\mathrm{ff}(\xe, \te, Z_i)\,,    \label{eq:kbr}
\end{equation}
where $\lambda_\mathrm{e} = 2\pi/m_\mathrm{e}$ is the electron Compton wavelength, the sum runs over ion species ($\mathrm{H}^+$, $\mathrm{He}^+$, $\mathrm{He}^{2+}$) with number densities $n_i$, and $g_\mathrm{ff}$ is the velocity-averaged free-free Gaunt factor---a quantum-mechanical correction to the classical Coulomb cross section that accounts for the finite de Broglie wavelength of the electron, approaching unity in the Born limit and growing logarithmically at low frequencies.
We use the softplus interpolation of \mbox{\cite{2011piim.book.....D}} for $g_{\rm ff}$.
The $\te^{-7/2}$ factor arises from thermally averaging the non-relativistic BR cross section over the Maxwell--Boltzmann distribution~\cite{Rybicki1979}.
The weaker temperature scaling of $K_\br$ relative to $K_\dc$ means BR dominates over DC at $z \lesssim 10^5$.

\subsection{Electron temperature}
\label{app:te_eq}
The electron temperature $\Te$ mediates the coupling between all three scattering processes: it determines the Compton drift rate and the DC/BR equilibrium target, and itself evolves under adiabatic expansion cooling.
Following Ref.~\cite{Chluba2012cosmotherm}, four effects compete to set $\Te$: (i) Compton scattering drives $\Te$ toward a value determined by the photon spectrum; (ii) external heating pushes $\Te$ above this equilibrium; (iii) DC/BR processes exchange energy between photons and electrons; and (iv) Hubble expansion cools the matter ($\Te \propto a^{-2}$) faster than the radiation ($\Tz \propto a^{-1}$).
The balance among these effects is governed by the evolution equation for $\re \equiv \Te / \Tz$~\cite{Chluba2012cosmotherm}:
\begin{equation}
    \frac{d\re}{d\tau} = \underbrace{\frac{4\tilde{\rho}_\gamma}{\alpha_h}\!\left(\left[\rho_\mathrm{eq} - \re \right] + \delta{\re}_\mathrm{inj} - H_{\dc/\br}(\re)\right)}_{\text{Compton + heating + DC/BR}} - \underbrace{H\,t_\mathrm{C}\,\re\vphantom{\frac{4}{a}}}_{\text{expansion}}\,,    \label{eq:te_ode}
\end{equation}
where $\tau$ is the Thomson optical depth, $t_\mathrm{C} = 1/(n_\mathrm{e}\sigma_\mathrm{T})$ the Thomson time, and $H$ the Hubble rate.
The photon energy density in units of the electron rest mass is $\tilde{\rho}_\gamma = \kappa_\gamma\,\tz^4\,G_3$ (with $\kappa_\gamma = 8\pi/\lambda_\mathrm{e}^3 \approx 1.76\times 10^{30}\,\mathrm{cm}^{-3}$), the heat capacity per volume is $\alpha_h = \tfrac{3}{2}(n_\mathrm{e} + n_\mathrm{H} + n_\mathrm{He})$, accounting for all particle species sharing the thermal bath (electrons, hydrogen nuclei, helium nuclei)~\cite{Chluba2012cosmotherm}, and $\delta{\re}_\mathrm{inj}$ is the dimensionless injection parameter defined below (Eq.~\eqref{eq:te_params}). 

The first term groups all effects that act through Compton coupling: equilibration toward $\rho_\mathrm{eq}$, external heating ($\delta{\re}_\mathrm{inj}$), and DC/BR back-reaction ($H_{\dc/\br}$).
The Compton equilibrium ratio is~\cite{Chluba2012cosmotherm}
\begin{equation}
    \rho_\mathrm{eq} = \frac{I_4}{4\,G_3}\,,    \label{eq:rho_eq}
\end{equation}
which is the temperature ratio that Compton scattering alone would enforce; $I_4$ and $G_3$ are as defined in Eq.~\eqref{eq:moments}, evaluated on the full distribution $n = \npl + \dn$.
For a Planck spectrum ($\dn = 0$), $\rho_\mathrm{eq} = 1$ exactly, whereas a spectral distortion pulls $\rho_\mathrm{eq}$ away from unity.
The DC/BR back-reaction integral is~\cite{Chluba2012cosmotherm}
\begin{equation}
    H_{\dc/\br}(\re) = \frac{1}{4\,G_3\,\tz}\int\bigl[1 - n\!\left(e^{\xe} - 1\right)\bigr]\,K(x)\,dx\,,    \label{eq:hdcbr}
\end{equation}
where $\xe = x\phi$ is the frequency in electron-temperature units and $K = K_\dc + K_\br$.
This integral measures the departure of the photon field from DC/BR equilibrium at $\Te$: it vanishes when $n$ is a Planck spectrum at the electron temperature, since the detailed-balance factor $[1 - n(e^{\xe}-1)]$ is zero in that case.
When DC/BR absorb soft injected photons, the absorbed energy heats the electrons; conversely, net DC/BR emission cools them.
The second term is the adiabatic cooling from Hubble expansion: matter cools as $\Te \propto a^{-2}$ while radiation cools as $\Tz \propto a^{-1}$, so in the ratio $\re = \Te/\Tz$ the net effect is a cooling rate $-H\,t_\mathrm{C}\,\re$.

Dividing Eq.~\eqref{eq:te_ode} by the Compton rate $4\tilde{\rho}_\gamma/\alpha_h$ gives an equivalent form that makes the timescale hierarchy explicit~\cite{Chluba2012cosmotherm}:
\begin{equation}
    \frac{d\re}{d\tau'} = \left(\left[\rho_\mathrm{eq} -\re \right] + \delta{\re}_\mathrm{inj} - H_{\dc/\br}(\re)\right) - \Lambda\,\re\,,    \label{eq:te_balance}
\end{equation}
where $\tau' = (4\tilde{\rho}_\gamma/\alpha_h)\,\tau$ is the Compton time measured in units of the Compton relaxation rate, and
\begin{equation}
    \delta{\re}_\mathrm{inj} \equiv \frac{t_\mathrm{C}}{4\,\tz}\frac{d(\drho)}{dt}\,,\qquad \Lambda \equiv \frac{H\,t_\mathrm{C}\,\alpha_h}{4\,\tilde{\rho}_\gamma}    \label{eq:te_params}
\end{equation}
are the dimensionless injection and expansion parameters.
Pre-recombination, the Compton rate vastly exceeds the Hubble rate ($\Lambda \ll 1$), so $\re$ equilibrates quasi-instantaneously and Eq.~\eqref{eq:te_balance} reduces to an algebraic balance.
Post-recombination, $X_e$ drops and Compton coupling weakens.
The numerical solution is described in Appendix~\ref{app:te_numerics}.

\section{Numerical method details}
\label{app:numerics}
This appendix describes the numerical methods used to solve the coupled photon Boltzmann equation (Eq.~\eqref{eq:boltzmann}).
The main difficulty is stiffness: DC/BR emission rates diverge as $K/x^3$ at low frequencies, so that the dimensionless rate $\Delta\tau \cdot K/x^3$ can exceed $10^7$ at the lowest frequencies.
We first describe the operator discretization (Sec.~\ref{app:time_integration}), then the coupled Newton solver including electron temperature feedback (Sec.~\ref{app:te_numerics}), followed by the frequency grid (Sec.~\ref{app:grid}), adaptive stepping (Sec.~\ref{app:stepping}), energy conservation (Sec.~\ref{app:energy}), and convergence tests (Sec.~\ref{app:convergence}), before turning to code verification: convergence at the design order to manufactured solutions (Sec.~\ref{app:mms}) and analytic constraints that pin the Kompaneets formulation (Sec.~\ref{app:moments}).

\subsection{Operator discretization}
\label{app:time_integration}
\label{app:imex}

At each time step, we advance $\dn$ simultaneously under both Compton scattering and DC/BR emission.
Rather than treating the two processes sequentially, which would introduce splitting errors, we discretize both operators and combine them into a single residual system (described below), which is then solved by the coupled Newton iteration of Sec.~\ref{app:te_numerics}.

As described in Sec.~\ref{sec:boltzmann}, we evolve $\dn$ rather than $n$, using the Planck identity $d\npl/dx + \npl(1+\npl) = 0$ to cancel the $O(1)$ background analytically.
The Compton operator acting on the frequency profile of $\dn$ is then (see Eq.~\eqref{eq:flux}):
\begin{equation}
    \mathcal{L}_K(\dn) = \frac{\te}{x^2}\frac{\partial}{\partial x}\!\left\{ x^4 \left[ (\phi - 1)\,\npl(1+\npl) + \frac{\partial\dn}{\partial x} + \phi\,(2\npl + 1)\,\dn + \phi\,\dn^2 \right] \right\}\,.    \label{eq:compton_op}
\end{equation}
The $\dn^2$ term is negligible for small distortions but is retained in the solver for completeness.
We discretize Eq.~\eqref{eq:compton_op} on the frequency grid by evaluating the expression inside the outer derivative at half-points $x_{i+1/2}$, with $\partial\dn/\partial x$ approximated by centered differences and $\dn$ by the average of neighboring grid values.
The discretized operator,
\begin{equation}
    \mathcal{L}_{K,i} = \frac{1}{x_i^2\,\Delta x_i}\left(\mathcal{F}_{i+1/2} - \mathcal{F}_{i-1/2}\right)\,,\end{equation}
couples each grid point $i$ only to its neighbors $i \pm 1$.
Explicitly, the half-point evaluation is
\begin{alignat}{2}
    \mathcal{F}_{i+1/2} &= &&\te\,x_{i+1/2}^4\!\left[(\phi{-}1)\,\npl(1{+}\npl)\big|_{i+1/2} + \frac{\dn_{i+1} - \dn_i}{\Delta x_{i+1/2}} \right. \nonumber \\
    & &&+ \phi\,(2\npl{+}1)\big|_{i+1/2}\,\overline{\dn}_{i+1/2} + \phi\,\overline{\dn}_{i+1/2}^2 \bigg]\,,
    \label{eq:flux_discrete}
\end{alignat}
where $\overline{\dn}_{i+1/2} \equiv (\dn_i + \dn_{i+1})/2$.

To integrate $\partial\dn/\partial\tau = \mathcal{L}_K$ forward in time, we use an \emph{implicit} scheme, in which $\dn^{n+1}$ appears on both sides of the update equation and must be solved for, rather than an \emph{explicit} scheme where $\dn^{n+1}$ is computed directly from $\dn^n$.
Implicit methods allow much larger time steps, which significantly improves solver performance.
Specifically, we use Crank--Nicolson~\cite{CrankNicolson1947}, which averages $\mathcal{L}_K$ over the step:
\begin{equation}
    \dn_i^{n+1} - \dn_i^n = \frac{\Delta\tau}{2} \left[ \mathcal{L}_{K,i}(\dn^{n+1}, \phi^{n+1}) + \mathcal{L}_{K,i}(\dn^n, \phi_\mathrm{p}) \right]\,,    \label{eq:cn}
\end{equation}
where superscript $n$ is the time-step index and subscript $i$ is the grid-point index.
Here $\phi^{n+1} = 1/\re^{n+1}$, which the Newton iteration described in Appendix~\ref{app:te_numerics} solves for.
In the second term, $\phi_\mathrm{p} = 1/\rho_\mathrm{p}$, where $\rho_\mathrm{p}$ solves Eq.~\eqref{eq:te_be} with $H_{\dc/\br}$ linearized about $\re^n$.
Both terms also use $\te = \rho_\mathrm{p}\,\tz$ in the prefactor.
Crank--Nicolson is second-order accurate in time for $\dn$ and unconditionally stable.

While Crank--Nicolson works well for Compton scattering, DC/BR emission rates diverge as $K/x^3$ at low frequencies, reaching $K/x^3 \sim 10^6$ per Thomson time at $x = 10^{-4}$, $z = 10^6$---a rate that vastly exceeds $1/\Delta\tau$ for any practical step size, placing the system in an extremely stiff regime.
Applying Crank--Nicolson to such stiff terms produces spurious oscillations rather than the correct exponential decay to equilibrium.
We therefore use a backward-Euler scheme for DC/BR, which damps stiff modes in a single step.
In this scheme, the DC/BR update at grid point $i$ is
\begin{equation}
    \dn_i^{n+1} = \frac{\dn_i^n + R_i\bigl[1 - {\npl}_i(e^{{\xe}_{,i}} - 1)\bigr]}{1 + R_i\,(e^{{\xe}_{,i}} - 1)}\,,
    \quad R_i \equiv \frac{\Delta\tau\,K_i}{x_i^3}\,,    \label{eq:dcbr_be}
\end{equation}
where $K_i = K_{\dc,i} + K_{\br,i}$ is the total emission coefficient (Eqs.~\eqref{eq:kdc}--\eqref{eq:kbr}).
When $R_i \gg 1$, this correctly relaxes $\dn$ to the equilibrium value $1/(e^{{\xe}_{,i}} - 1) - {\npl}_i$, which vanishes when $\Te = \Tz$.

Rather than solving the Compton and DC/BR updates sequentially, we combine them into a single nonlinear system and solve for $\dn^{n+1}$ all at once.
We define the residual $\mathcal{R}_i$ as the amount by which a trial solution $\dn_i^{n+1}$ fails to satisfy the coupled update equations at each grid point $i$:
\begin{align}
    \mathcal{R}_i &= \dn_i^{n+1} - \dn_i^n - \frac{\Delta\tau}{2}\left[\mathcal{L}_{K,i}(\dn^{n+1}, \phi^{n+1}) + \mathcal{L}_{K,i}(\dn^n, \phi_\mathrm{p})\right] \nonumber \\
        &\quad - \Delta\tau\,\frac{K_i}{x_i^3}\left[1 - {\npl}_i(e^{{\xe}_{,i}} - 1) - \dn_i^{n+1}(e^{{\xe}_{,i}} - 1)\right]\,.
    \label{eq:residual}
\end{align}
The correct solution has $\mathcal{R}_i = 0$ for all $i$.
Because the DC/BR absorption term $\dn(e^{\xe} - 1)$ makes this system nonlinear, it must be solved iteratively; the Newton iteration that accomplishes this, including the coupled electron temperature feedback, is described next.

\subsection{Coupled Newton solver}
\label{app:te_numerics}

The $N$ photon residuals $\mathcal{R}_i$ (Eq.~\eqref{eq:residual}) depend on the electron temperature through $\phi = 1/\re$ in the Kompaneets operator and through the DC/BR equilibrium target.
If $\re$ is held fixed during the solve, the solver cannot self-consistently capture the feedback between the photon spectrum and the electron temperature: any process that transfers energy between photons and electrons---DC/BR emission and absorption, external heating---should feed back into the Kompaneets operator within the same time step.
We therefore promote $\re$ to an $(N\!+\!1)$-th unknown, solved simultaneously with $\dn$.
We first describe the backward-Euler discretization of the $\re$ equation, then assemble the full $(N\!+\!1)$-dimensional bordered tridiagonal system and its $O(N)$ solution.

\paragraph{Electron temperature discretization.}
The electron temperature ratio $\re$ is advanced by backward Euler applied to Eq.~\eqref{eq:te_balance}.
Writing $\re^n$ for the value at the start of the step and $\Delta\tau \equiv (4\tilde{\rho}_\gamma/\alpha_h)\,\Delta\tau_\mathrm{Th}$ for the rescaled Compton relaxation step (where $\Delta\tau_\mathrm{Th}$ is the Thomson optical depth traversed in one step):
\begin{equation}
    \re^{n+1} = \re^n + \Delta\tau\left[(\rho_\mathrm{eq} - \re^{n+1}) + \delta{\re}_\mathrm{inj} - H_{\dc/\br}(\re^{n+1}) - \Lambda\,\re^{n+1}\right].    \label{eq:te_be}
\end{equation}
Linearizing $H_{\dc/\br}$ around $\re^n$ and solving for $\re^{n+1}$ gives:
\begin{equation}
    \re^{n+1} = \frac{\re^n + \Delta\tau\left(\rho_\mathrm{eq} + \delta{\re}_\mathrm{inj} - H_{\dc/\br}^n + H'^n\,\re^n\right)}{1 + \Delta\tau\left(1 + H'^n + \Lambda\right)}\,,    \label{eq:te_be_solve}
\end{equation}
where $H_{\dc/\br}^n \equiv H_{\dc/\br}(\re^n)$ and $H'^n \equiv \partial_{\re} H_{\dc/\br}|_{\re^n}$, computed analytically by differentiating the integrand of Eq.~\eqref{eq:hdcbr} at fixed $\dn$: the Bose factor, the temperature dependence of the bremsstrahlung coefficient, and the Gaunt factor each contribute in closed form.

This formulation has two important limits.
Pre-recombination ($\Delta\tau \gg 1$), the leading $\re^n$ and the 1 in the denominator are negligible, and Eq.~\eqref{eq:te_be_solve} reduces to the quasi-stationary solution $\re \approx (\rho_\mathrm{eq} + \delta{\re}_\mathrm{inj} - H_{\dc/\br}^n + H'^n\re^n)/(1 + H'^n + \Lambda)$.
After baryons decouple from the CMB, Compton coupling freezes out and $\re$ evolves only through adiabatic cooling ($\re \propto 1+z$).

Direct evaluation of $\rho_\mathrm{eq} = I_4/(4 G_3)$ from $n = \npl + \dn$ suffers from catastrophic cancellation, since the $O(1)$ Planck contributions to $I_4$ and $G_3$ nearly cancel against the denominator to leave the small distortion signal.
We avoid this by expanding to first order in $\dn$~\cite{Chluba2012cosmotherm}:
\begin{equation}
    \rho_\mathrm{eq} - 1 \approx \frac{\Delta I_4}{4\,\Gpl{3}} - \frac{\Delta G_3}{\Gpl{3}}\,,
    \label{eq:rho_e_pert}\end{equation}
where $\Delta I_4 = \int x^4 (2\npl+1)\,\dn\,dx$ and $\Delta G_3 = \int x^3\,\dn\,dx$ are the perturbations of the fourth and third spectral moments.
Both numerator integrals involve only $\dn$, eliminating the cancellation.

\paragraph{Bordered tridiagonal system.}
The goal is to find $\dn^{n+1}$ and $\re^{n+1}$ that simultaneously satisfy the implicit photon update (Eq.~\eqref{eq:residual}, $\mathcal{R}_i = 0$) and the implicit $\re$ update (Eq.~\eqref{eq:te_be}).
Combining these into a single $(N\!+\!1)$-dimensional system and solving by Newton iteration~\cite{Press2007}, the correction $(\delta\dn,\,\delta\re)$ at each step is
\begin{equation}
    \left(\!\begin{array}{c|c} T & \mathbf{c} \\[2pt] \hline \\[-8pt] \mathbf{b}^\top & d \end{array}\!\right)
    \begin{pmatrix} \delta\dn \\ \delta\re \end{pmatrix}
    = -\begin{pmatrix} \mathbf{r} \\ r_\rho \end{pmatrix}\!,
    \label{eq:bordered}
\end{equation}
where $\mathbf{r} = (\mathcal{R}_1, \ldots, \mathcal{R}_N)$ collects the photon residuals from Eq.~\eqref{eq:residual} at each grid point, and $r_\rho \equiv \re^{n+1} - \text{RHS of Eq.~\eqref{eq:te_be}}$ is the mismatch of the backward-Euler $\re$ update. The solution $(\dn^{n+1}, \re^{n+1})$ satisfies the coupled implicit system when $\mathbf{r} = 0$ and $r_\rho = 0$; Newton iteration finds this root.

The Jacobian has a \emph{bordered tridiagonal} structure, reflecting the physics of the coupling.
The upper-left block $T$ is the $N\times N$ tridiagonal Jacobian $\partial\mathcal{R}_i/\partial\dn_j$: the Kompaneets operator couples each grid point only to its nearest neighbors ($i \pm 1$), while DC/BR acts locally on the diagonal.

The off-diagonal blocks encode the photon--electron coupling.
The column $\mathbf{c} = \partial\mathcal{R}_i/\partial\re$ captures how a change in $\re$ affects the photon equation: since $\phi = 1/\re$, the Kompaneets drift rate at each frequency shifts, giving $c_i \propto -\re^{-2}\,x^4\,n(1\!+\!n)$.
The row $\mathbf{b}^\top = \partial r_\rho/\partial\dn_j$ captures the reverse: how changes in the photon spectrum affect $\Te$ through the DC/BR back-reaction integral $H_{\dc/\br}$ (Eq.~\eqref{eq:hdcbr}).
The scalar $d = \partial r_\rho/\partial\re = 1 + \Delta\tau(1 + H' + \Lambda)$ is always large ($\Delta\tau \gg 1$ pre-recombination), which ensures that $\re$ responds stably to perturbations.

This bordered tridiagonal system is solved in $O(N)$ operations.
We first obtain two auxiliary solutions from the standard Thomas algorithm applied to the tridiagonal block $T$:
\begin{equation}
    T\,\mathbf{u} = -\mathbf{r}, \qquad T\,\mathbf{v} = \mathbf{c}\,.
    \label{eq:thomas_pair}
\end{equation}
Here $\mathbf{u}$ is the Newton correction that $\dn$ \emph{would} receive if $\re$ were frozen, and $\mathbf{v}$ is the sensitivity of $\dn$ to a unit change in $\re$.
Substituting into the bottom row and solving for $\delta\re$ gives
\begin{equation}
    \delta\re = \frac{-r_\rho - \mathbf{b}^\top\!\mathbf{u}}{d - \mathbf{b}^\top\!\mathbf{v}}\,,\qquad
    \delta\dn = \mathbf{u} - \mathbf{v}\,\delta\re\,.
    \label{eq:bordered_solve}
\end{equation}
The denominator $d - \mathbf{b}^\top\!\mathbf{v}$ is always well-separated from zero because the diagonal dominance of $d \sim \Delta\tau \gg 1$ far exceeds the off-diagonal coupling $\mathbf{b}^\top\!\mathbf{v}$.
At each Newton step, the two Thomas solves and the scalar correction replace a single Thomas solve, and the iteration continues until $\max(|\delta\dn|, |\delta\re|)$ falls below the convergence tolerance $\epsilon = 10^{-8} \max_i |\dn_i| + 10^{-14}$.
Convergence typically requires 2--4 iterations.

This coupling ensures that energy transferred between photons and electrons is self-consistently routed through $\re$ into Kompaneets redistribution within each Newton iteration.
The full bordered system is solved at every time step; for pure heat injection the bordered correction $\mathbf{v}\,\delta\re$ is naturally small and the cost is negligible.

\subsection{Frequency grid}
\label{app:grid}

The distortion $\dn(x)$ spans five orders of magnitude in dimensionless frequency $x = 2\pi \nu/\Tz$, from $x \sim 10^{-4}$ (where DC/BR emission diverges as $1/x^3$) to $x \sim 50$.
A uniform grid would waste resolution---the DC/BR rates require dense sampling at low $x$, while the energy integrals (e.g., $\int x^3 \npl\,dx = \pi^4/15$) require coverage to high $x$.
We therefore use a composite grid~\cite{Chluba2012cosmotherm}: logarithmically spaced at $x < x_t$ (capturing the DC/BR divergence) transitioning to linearly spaced at $x > x_t$ (efficiently covering the Wien tail).

The Python interface, the Rust library, and the command-line interface all use the same grid by default: $N = 4000$, $x_\mathrm{min} = 10^{-5}$, $x_\mathrm{max} = 60$, $x_t = 0.5$, with $35\%$ of points in the logarithmic region, i.e. where $x<x_t$.
The low $x_\mathrm{min}$ resolves DC/BR thermalization at $z_h \gtrsim 10^6$, where it extends to lower frequencies.
At both boundaries we impose $\dn = 0$: $x_\mathrm{max} \gtrsim 30$ is chosen to lie well above the largest signal frequency of interest (where $\npl \sim e^{-x_\mathrm{max}}$ is exponentially suppressed), and at $x_\mathrm{min}$ the DC/BR rates drive $\dn$ to its equilibrium value on timescales much shorter than a single step, so the boundary value is never dynamically important.

For photon injection at low frequencies ($x_\mathrm{inj} \lesssim 0.1$), the solver automatically inserts a refinement zone of $\sim 300$ points within $\pm 5\sigma_x$ of $x_\mathrm{inj}$ to resolve the sharp spectral feature and DC/BR absorption, achieving convergence already on an $N = 2000$ grid.
The grid can be extended up to $x_\mathrm{max} \sim 150$ for high-frequency photon injection; we have not extensively tested photon injection beyond this point.

\subsection{Adaptive redshift stepping}
\label{app:stepping}

The integration proceeds from high to low redshift, with $\Delta z$ adapted to the local physics.
Two criteria determine the step size.

The first is the \emph{Thomson optical depth limit}: $\Delta\tau < \Delta\tau_\mathrm{max} = 10$.
This ensures that the implicit time-stepping remains accurate (even though it is unconditionally stable, accuracy degrades for very large steps).

The second is the \emph{Compton evolution limit}: $\Delta z = f_\mathrm{tol} \times t_C H(1+z)/\te$, where $f_\mathrm{tol} = 0.02$ and $t_C = 1/(n_e \sigma_T)$ is the Thomson time.
The ratio $t_C H / \te$ measures how many Hubble times it takes for Compton scattering to redistribute energy by one $e$-folding.
When this ratio is small (at high redshift, where Compton scattering is fast), the distortion evolves rapidly and needs small steps to track the evolution accurately; when it is large (at low redshift, where scattering freezes out), larger steps suffice.
The prefactor $f_\mathrm{tol} = 0.02$ was calibrated by convergence testing: halving it changes $\mu$ and $y$ by $< 0.2\%$.

When a single-burst injection is enabled, the step is further limited to $\sigma_z / 10$ (where $\sigma_z$ is the temporal width of the injection profile) to resolve the source.
Below $z \approx 1575$, where $X_\mathrm{e}$ depends more sensitively on $\Te$, the step is also capped at $\Delta z = 0.005\,z$ (Sec.~\ref{sec:boltzmann}).
In the Python interface, integration runs from $z_h + 7\sigma_z$ (burst) or $z \sim 3 \times 10^6$ (continuous) down to $z = 10$ by default; the Rust library and the command-line interface stop at the same redshift.
The number of steps depends strongly on injection redshift: $\sim 10^3$ in the $y$-era ($z_h \sim 10^3$), most of them from the $0.005\,z$ cap below $z \approx 1575$, to $\sim 10^4$ for a $\mu$-era burst at $z_h = 5\times10^5$ and $\sim 6\times10^4$ for continuous injection from $z = 3\times10^6$, with most steps concentrated at high redshifts where the Compton time is very short.

\subsection{Energy conservation and temperature-shift subtraction}
\label{app:energy}
The solver does not enforce energy conservation as a hard constraint on $\dn$.
Instead, energy is conserved to sub-percent accuracy through two properties of the numerical scheme: (i) the implicit time-stepping conserves energy to high accuracy by construction, and (ii) a number-conserving temperature-shift subtraction removes the unobservable $\Gbb$ component that accumulates from DC/BR photon creation, preventing it from feeding back into $\re$ and degrading energy balance.
We discuss each in turn.

\begin{figure}[!t]
\centering
\includegraphics[width=0.48\textwidth]{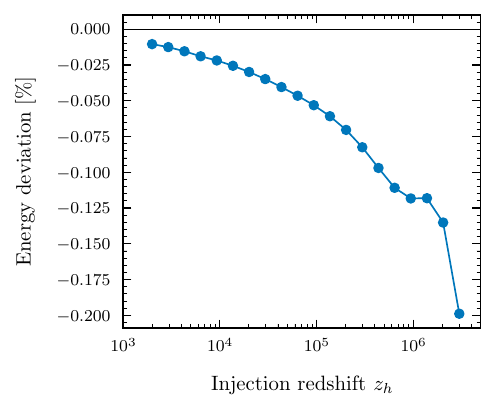}\hfill
\includegraphics[width=0.48\textwidth]{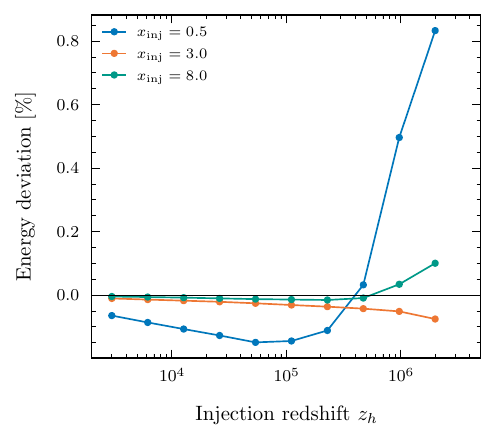}
\caption{Energy conservation diagnostic: percentage deviation of recovered from injected energy, $(\Delta\rho_\mathrm{PDE} - \Delta\rho_\mathrm{inj}) / \Delta\rho_\mathrm{inj}$, as a function of injection redshift.
\emph{Left:} Single-burst heat injection with $\drho = 10^{-5}$. Both panels use $N = 8000$ grid points and $\Delta\tau_\mathrm{max} = 3$.
\emph{Right:} Monochromatic photon injection with $\Delta N/N = 10^{-5}$ at three injection frequencies.
~\notebooklink{notebooks/paper\_figures/energy\_conservation.ipynb}}
\label{fig:energy}
\end{figure}

Neither the Crank--Nicolson scheme (for Compton) nor the backward-Euler scheme (for DC/BR) conserves energy exactly.
The solver computes the Compton equilibrium temperature $\rho_\mathrm{eq}$ from the spectrum at the start of each step, and the Crank--Nicolson update uses an estimate of $\re$ that differs slightly from the converged value $\re^{n+1}$. This error accumulates over the ${\sim}\,10^4$ steps of a $\mu$-era run.
Figure~\ref{fig:energy} shows the energy diagnostic $(\Delta\rho_\mathrm{PDE} - \Delta\rho_\mathrm{inj})/\Delta\rho_\mathrm{inj}$, where the total recovered energy $\Delta\rho_\mathrm{PDE}$ includes both the distortion $\int x^3\,\dn\,dx/\Gpl{3}$ and the accumulated temperature shift $4\,\delta T/T$ removed by the number-conserving subtraction.
For heat injection (panel~a), the recovered energy falls short by $0.01\%$ at the low-$z$ end. The deficit grows with $z_h$ to $0.2\%$ at $z_h = 3 \times 10^6$, with a plateau near $0.11\%$ around $z_h \sim 10^6$.
For photon injection (panel~b), energy conservation is sub-percent for $x_\mathrm{inj} \geq 3$ across all redshifts.
For soft photon injection ($x_\mathrm{inj} = 0.5$), the deviation grows to $0.5\%$ at $z_h \approx 10^6$ and $0.85\%$ at $z_h = 2 \times 10^6$.
At higher redshifts ($z_h \gtrsim 3 \times 10^6$), thermalization is nearly complete ($\Jbb \lesssim 0.06$), so the observable distortion is exponentially suppressed, and FIRAS/PIXIE constraints on photon injection are driven by lower redshifts where the conservation is sub-percent.

DC/BR processes continuously create photons during the thermalization process, producing an accumulating temperature shift in $\dn$.
Since FIRAS measures the CMB with $T_0$ as a free parameter, this temperature shift is unobservable~\cite{Chluba2012cosmotherm}.
Following \cosmotherm{}, we subtract it after each step (at $z > 5 \times 10^4$) by projecting out the number-violating component, $\int x^2 \dn\,dx = 0$:
\begin{equation}
    \dn \leftarrow \dn - \frac{\int x^2 \dn\,dx}{\int x^2\,\Gbb(x)\,dx}\,\Gbb(x)\,.
\end{equation}
It is orthogonal to the observable distortion because both $M(x)$ and $\Ysz(x)$ conserve photon number ($\int x^2 M\,dx = 0$, $\int x^2 \Ysz\,dx = 0$).
The subtraction also improves numerical stability at high redshift by preventing the accumulated temperature shift from feeding back into $\re$.

\subsection{Decomposition of \texorpdfstring{$\dn(x)$}{Δn(x)} into \texorpdfstring{$\mu$, $y$, $\Delta T/T$}{μ, y, ΔT/T}}
\label{app:decomposition}

The solver produces a numerical $\dn(x)$ that is a generic mixture of temperature-shift, $\mu$-, and $y$-type distortions, plus a residual component not well-described by any of these templates.
Extracting $\mu$, $y$, and $\Delta T/T$ from this spectrum requires a choice of fitting procedure.

A linear least-squares fit of $\dn$ to $\{\Gbb, M, \Ysz\}$ has a fundamental problem: the temperature-shift and linearized-$\mu$ shapes are nearly the same function.
Linearizing a small chemical potential gives $[e^{x+\mu}-1]^{-1} - \npl(x) \approx -\mu\,\Gbb(x)/x$, while a small temperature shift gives $\npl(x/(1+\delta)) - \npl(x) \approx \delta\,\Gbb(x)$.
Both contributions carry the same $\Gbb(x)$ envelope and differ only by the slowly varying factor $1/x$, so across the FIRAS band $x \in [0.5, 10]$ a positive $\mu$ and a negative $\delta$ produce almost identical spectral imprints---any fitted $\mu$ can be partly reabsorbed into the fitted temperature and vice versa.

Therefore, to avoid this problem, we adopt the formulation of Ref.~\cite{BianchiniFabbian2022}, in which the chemical potential is kept \emph{inside} the Bose--Einstein exponential rather than linearized. The spectral distortion is then written as
\begin{equation}
    \dn_\mathrm{model}(x;\,\mu,\delta,y) =
      \left[\frac{1}{e^{x+\mu} - 1} - \npl(x)\right]
    + \delta\,\Gbb(x)
    + y\,\Ysz(x)\,,    \label{eq:bf_model}
\end{equation}
with $\delta \equiv \Delta T/\Tz$.
We fit $\dn$ to Eq.~\eqref{eq:bf_model} by nonlinear least squares on the band $x \in [0.5, 18]$ ($\nu \in [28, 1020]\,\mathrm{GHz}$ at $T_0 = 2.725\,\mathrm{K}$), minimizing the intensity residual $\int [x^3(\dn - \dn_\mathrm{model})]^2\,dx$.
The extracted $\mu$ is then the physical Bose--Einstein chemical potential rather than the amplitude of $M$, which is the convention used in direct analyses of absolute-spectrum data~\cite{Fixsen1996, BianchiniFabbian2022}.

We also implement an alternate fitting procedure from Ref.~\cite{CJ2014} in \spectroxide{}; the two methods give identical results on $\mu$ and $y$ across the entire \cosmotherm{} Green's function table.

\subsection{Convergence tests}
\label{app:convergence}

\begin{figure}[!t]
\centering
\includegraphics[width=\textwidth]{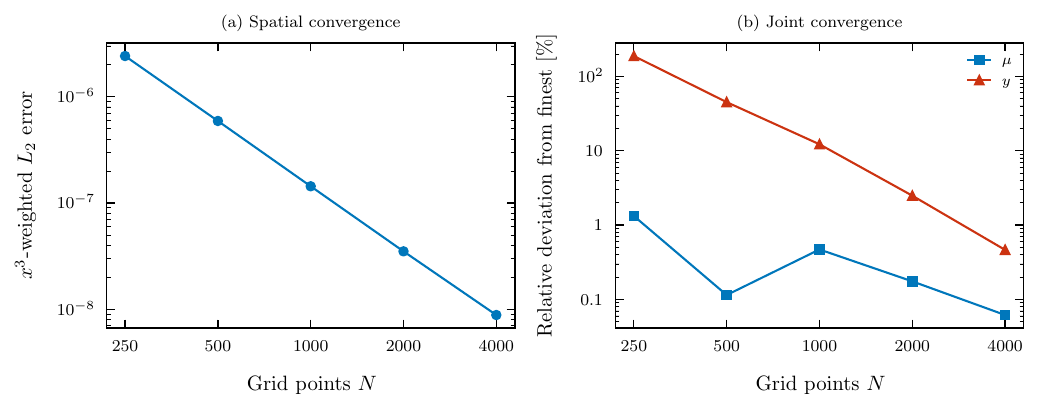}
\caption{Convergence of the IMEX solver.
(a)~Spectral $L_2$ error for a Gaussian initial condition evolved under pure Kompaneets scattering, showing the expected $O(N^{-2})$ spatial convergence of the Crank--Nicolson scheme.
(b)~Relative deviation of $\mu$ and $y$ from the finest-resolution run ($N = 8000$, $f_\mathrm{tol} = 5 \times 10^{-4}$) under simultaneous grid and time-step refinement, for a single-burst heat injection at $z_h = 2 \times 10^5$ with $\drho = 10^{-5}$.
~\notebooklink{notebooks/paper\_figures/convergence\_study.ipynb}}
\label{fig:convergence}
\end{figure}

We verify numerical convergence by varying the grid resolution, time-step size, and the treatment of DC/BR coupling.
Figure~\ref{fig:convergence} demonstrates convergence under simultaneous refinement of the grid and time step.
Panel~(a) isolates the spatial discretization of the Kompaneets operator by evolving a smooth Gaussian perturbation $\dn(x) \propto \exp[-(x-3)^2/(2\sigma^2)]$ with $\sigma = 0.5$ under pure Compton scattering with no source term or DC/BR.
This is the cleanest convergence test because the error comes entirely from the Crank--Nicolson spatial discretization.
The $x^3$-weighted spectral $L_2$ error $\|\dn_N - \dn_\mathrm{ref}\|_{L_2} \equiv [\int x^3\,(\dn_N - \dn_\mathrm{ref})^2\,dx]^{1/2}$ (where $\dn_\mathrm{ref}$ is the $N = 8000$ solution, with coarser grids interpolated onto the reference) converges at the expected $O(N^{-2})$ rate.
Panel~(b) tests the full coupled system (Compton + DC/BR + $\Te$ feedback) with a single-burst heat injection at $z_h = 2 \times 10^5$ with $\drho = 10^{-5}$ (width $\sigma_z = 3000$, integrated from $z = 5 \times 10^5$ to $10^4$ with $\Delta\tau_\mathrm{max} = 200$), showing the relative deviation of $\mu$ and $y$ from the $N = 8000$ reference as both $N$ and $f_\mathrm{tol}$ (the Compton evolution time-step prefactor) are refined simultaneously.
At $N = 2000$, $\mu$ lies within $0.2\%$ of the reference and $y$ within $2.5\%$; at $N = 4000$, within $0.06\%$ and $0.5\%$. At this redshift $y$ is about $10^{-3}$ of $\mu$.

\subsection{Verification with manufactured solutions}
\label{app:mms}

The convergence tests above measure self-convergence and demonstrate solver stability, but they are not sensitive to a mistake in the differential equations themselves.
The method of manufactured solutions (MMS)~\cite{Roache2002} addresses this by reversing the usual direction of a test problem: rather than searching for a scenario with a known analytic solution, we pick a solution first and construct the problem it solves.
Substituting a chosen trial distortion $\dn_\mathrm{m}(x,\tau)$ into the evolution equation leaves a residual, $S = \partial_\tau \dn_\mathrm{m} - \mathcal{L}[\dn_\mathrm{m}]$.
Injected back as a source term, $S$ makes $\dn_\mathrm{m}$ an exact solution of the sourced equation.

We therefore run the solver with $S$ supplied as a source term, and compare its output against $\dn_\mathrm{m}$.
For the tests here, the trial solution is a Gaussian bump with a slowly oscillating amplitude: $\dn_\mathrm{m} = a(\tau)\,g(x)$ with $g(x) = e^{-(x-x_0)^2/2\sigma^2}$ ($x_0 = 3$, $\sigma = 0.6$) and $a(\tau) = 10^{-3}\bigl[1 + \tfrac{1}{2}\sin(2\pi\tau/50)\bigr]$. $S$ can then be computed analytically. 

We then repeat the run on a ladder of successively finer frequency grids and time steps, and measure the rate at which the numerical solution approaches $\dn_\mathrm{m}$.
A discretization bug appears as a wrong convergence order or a wrong limit; what MMS certifies is the discretization of the operator as coded, not the physical correctness of the operator itself.

\begin{figure}[!t]
\centering
\includegraphics[width=\textwidth]{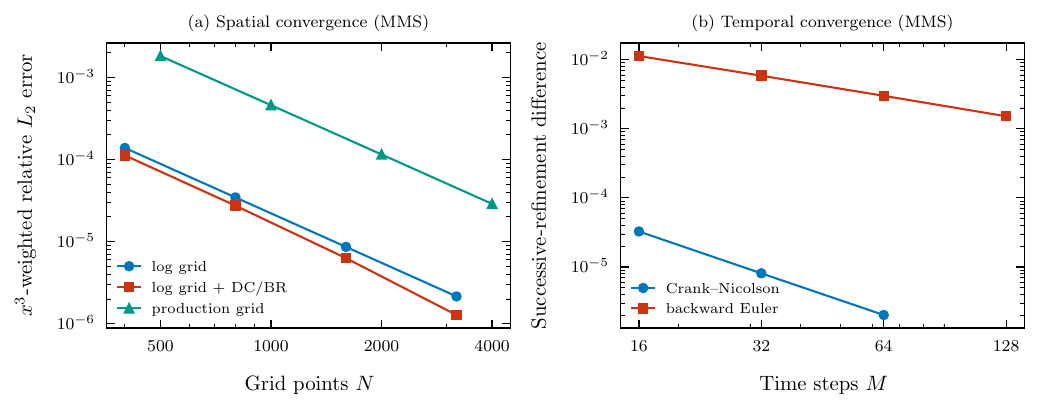}
\caption{Convergence to manufactured solutions.
(a)~Spatial error on three grid configurations.
The Crank--Nicolson discretization is second order, so the expected log--log slope is $-2$ (error $\propto N^{-2}$); all three configurations follow it.
(b)~Temporal convergence via successive step halvings.
The expected slopes are $-2$ for the Crank--Nicolson Compton branch and $-1$ for the backward-Euler DC/BR branch, and both branches follow them.
~\notebooklink{notebooks/paper\_figures/mms\_convergence.ipynb}}
\label{fig:mms}
\end{figure}

We run MMS in several configurations of increasing completeness, because the isolated tests and the end-to-end test answer different questions: an operator tested alone yields a clean order measurement and localizes any failure to a specific scheme, while the assembled pipeline can only show that composing the pieces introduces no new error.
Figure~\ref{fig:mms} shows the isolated tests.
We test the Kompaneets operator both alone and together with a stiff absorption term that mimics the steep $1/x^3$ frequency dependence of DC/BR emission, the most numerically difficult term in the problem.
Panel~(a) refines the frequency grid at fixed time step: the error falls as $N^{-2}$ in every configuration, as expected for the second-order Crank--Nicolson discretization.
Panel~(b) refines the time step at fixed grid: the Compton update converges at second order and the DC/BR-like term at first order, matching the design orders of the two schemes (Sec.~\ref{app:imex}).

The tests above exercise the Compton and DC/BR updates in isolation.
Two further checks address the rest of the pipeline.
The first is an end-to-end MMS run through the full solver, with adaptive redshift stepping and the recombination history active; it reproduces its manufactured solution to $2.5\times10^{-4}$ in the relative $x^3$-weighted $L_2$ norm of Fig.~\ref{fig:mms}.
We also check that photon number is conserved and find that the number integral $\int x^2\,\dn\,dx$ is unchanged to machine precision over 200 steps.

\subsection{Analytic checks of the Kompaneets operator}
\label{app:moments}
In order to further verify the correct implementation of the Kompaneets equation beyond ensuring its correct discretization, we perform analytic checks using relations derived by hand from the continuum Kompaneets equation.
Two properties of the operator serve: a hierarchy of moment equations, and its exact linear response to a small electron--photon temperature difference.
We derive the hierarchy first.
Multiplying Eq.~\eqref{eq:kompaneets} by $x^k$ and integrating over frequency converts it into a hierarchy of ordinary differential equations for the moments $\Delta G_k \equiv \int x^k\,\dn\,dx$; at $\phi = 1$,
\begin{equation}
    \frac{d\,\Delta G_k}{dy_C} = (k-2)(k+1)\,\Delta G_k - (k-2)\bigl[\Delta G_{k+1} + C_k\bigr]\,,
    \qquad
    C_k = \int x^{k+1}\bigl(2\npl\,\dn + \dn^2\bigr)\,dx\,,
    \label{eq:moment_hierarchy}
\end{equation}
where $dy_C = \te\,d\tau$ is the Compton time variable and the distortion is assumed to vanish at the grid edges.
The low moments describe familiar physics: $k = 2$ expresses photon-number conservation, and $k = 3$ contains the Zeldovich--Sunyaev energy-exchange law~\cite{Zeldovich1969}, $d\ln \Delta G_3/dy_C = 4 - x_0$ for a narrow dilute line at $x_0$---Doppler boosting against recoil.
The test is then direct: evolve a simple spectrum, measure each $d\,\Delta G_k/dy_C$ numerically, and demand that it obey Eq.~\eqref{eq:moment_hierarchy}.

Each check pairs an initial spectrum with the property it tests.
The first is the hierarchy itself, whose coefficients are fixed by the drift and recoil terms of the $\phi = 1$ flux.
We initialize a narrow Gaussian line, $\dn = A\,e^{-(x-x_0)^2/2\sigma^2}$ with $x_0 = 7$, $\sigma = 0.5$, and $A = 10^{-3}$, evolve it under pure Compton scattering, and measure each rate $d\,\Delta G_k/dy_C$ by finite differences at $k = 3$, $4$, $5$ simultaneously.
For each $k$, the measured and predicted $d\Delta G_k/dy_C$ agree to a relative accuracy of $2\times10^{-4}$.
The second check is the linear response: the $(\phi-1)$ flux branch converts heating into a $y$-distortion, a channel every $\phi = 1$ test leaves untouched.
Here we start from an unperturbed blackbody and hold $\Te$ above $\Tz$ by one part in $10^3$: the distortion that builds up must be a $y$-distortion, $\dn = \Delta y\,\Ysz(x)$, with the standard Compton-$y$ amplitude $\Delta y = \int (\te - \tz)\,d\tau$.
The code matches this prediction pointwise to $3.4\times10^{-4}$ of the distortion's peak.

\subsection{Performance}
\label{app:performance}

Table~\ref{tab:performance} summarizes representative timings.
Individual PDE runs are single-threaded; the multi-redshift sweep exploits Rust's safe concurrency primitives to parallelize across injection redshifts with no additional infrastructure.
Single-burst runtime depends strongly on injection redshift, from sub-second in the $y$-era to $\sim$22\,s for deep thermalization ($z_h = 3\times10^6$). Continuous injection scenarios (decaying particles, DM annihilation) integrate from $z \sim 3\times10^6$ and take $\sim$19\,s. Monochromatic photon injection, which requires fine stepping through the injection window, completes in $\sim$19\,s.
The package also includes a pure-Python analytic Green's function (Table~\ref{tab:performance}, bottom rows) for fast approximate calculations; the Python interface is described in Appendix~\ref{app:python}.

\begin{table}[!tbp]
\centering
\caption{Representative wall-clock times measured via the Python wrapper on an AMD Ryzen 5 8645HS (6 cores / 12 threads, 7~GB RAM).
PDE runs are single-threaded except where noted; all use the Python wrapper's default settings ($N = 4000$ grid points, production-quality time-step control).
Single-burst runs start at $z_\mathrm{start} = z_h + 7\sigma_z$; continuous injection runs start at $z_\mathrm{start} = 3 \times 10^6$ (energy injected above $z_\mathrm{th} \approx 1.98 \times 10^6$ is fully thermalized).
The first invocation adds $\sim 15$~s for Rust binary compilation, which is cached thereafter; the times listed are for subsequent (cached) calls.
}
\label{tab:performance}
\begin{tabular}{lc}
\toprule
Operation & Time \\
\midrule
PDE single burst ($y$-era, $z_h = 10^3$) & \qty{0.04}{\second} \\
PDE single burst ($\mu$--$y$ transition, $z_h = 10^5$) & \qty{0.7}{\second} \\
PDE single burst (thermalization, $z_h = 3 \times 10^6$) & \qty{22}{\second} \\
PDE decaying particle & \qty{19}{\second} \\
PDE $s$-wave annihilation & \qty{18}{\second} \\
PDE monochromatic photon injection ($x_\mathrm{inj} = 0.01$, $z_h = 5 \times 10^5$) & \qty{19}{\second} \\
15-point sweep (parallel) & \qty{27}{\second} \\
\midrule
GF single spectrum (1000 pts) & \qty{20}{\milli\second} \\
GF convolution ($1000 \times 1000$) & \qty{71}{\milli\second} \\
GF $\mu + y$ from heating & \qty{0.12}{\milli\second} \\
\bottomrule
\end{tabular}
\end{table}

\section{Code architecture and usage}
\label{app:code}

\subsection{Design philosophy}
\label{app:design}
Rust provides compiled performance comparable to C/Fortran, memory safety without a garbage collector, and a built-in package manager and test framework that simplifies development.

\paragraph{Zero dependencies.}
The Rust code uses only the standard library (dev-dependencies: \code{approx} for float comparison and \code{criterion} for benchmarking), so it builds on any platform with a Rust compiler and every line can be audited directly.

\paragraph{Module structure.}
The Rust codebase is organized so that each physical process has its own module, with equations documented in comments and variable names matching the notation of Sec.~\ref{sec:physics}:
\begin{itemize}
    \item \code{kompaneets.rs} --- Compton scattering via the Fokker--Planck equation (Sec.~\ref{sec:boltzmann}). Contains the IMEX solver: Crank--Nicolson for the Compton operator, backward Euler for DC/BR, with nonlinear Newton iteration. This is the largest and most numerically delicate module.
    \item \code{double\_compton.rs} --- DC emission ($\gamma e \to \gamma\gamma e$). Implements Eq.~\eqref{eq:kdc} with the relativistic correction of Ref.~\cite{Chluba2007dc}.
    \item \code{bremsstrahlung.rs} --- BR emission ($e + X^{+Z} \to e + X^{+Z} + \gamma$). Implements Eq.~\eqref{eq:kbr} with the Gaunt factor interpolation of Ref.~\cite{2011piim.book.....D}.
    \item \code{electron\_temp.rs} --- Electron temperature $\Te$. Helper functions for the Compton equilibrium ratio $\rho_\mathrm{eq}$ (Eq.~\eqref{eq:rho_eq}); the perturbative (Eq.~\eqref{eq:rho_e_pert}) and exact evaluation modes, along with the full backward-Euler ODE integration (Eq.~\eqref{eq:te_ode}), are in \code{solver.rs}.
    \item \code{recombination.rs} --- Ionization fraction $X_\mathrm{e}(z)$. Peebles three-level atom~\cite{Peebles1968} for hydrogen at $z < 1575$ with the RECFAST corrections~\cite{Seager1999,Rubino-Martin2010}, advanced alongside $\Te$ by \code{solver.rs}, Saha equilibrium above; helium uses Saha equilibrium at all redshifts.
    \item \code{dark\_photon.rs} --- Resonant photon--dark-photon conversion in the narrow-width approximation (Sec.~\ref{sec:dark_photon}).
    \item \code{solver.rs} --- Main PDE integrator coupling all processes with adaptive redshift stepping.
    \item \code{energy\_injection.rs} --- All injection scenarios from Secs.~\ref{sec:heat_scenarios} and~\ref{sec:photon_scenarios}.
\end{itemize}

\paragraph{Test-driven development.}
Over 950 automated tests across Rust and Python (unit, integration, and doc-tests) serve as executable documentation and catch regressions.
The test suite is organized hierarchically: mathematical identities, analytic limits, PDE convergence, physical scenarios, and literature benchmarks.

\subsection{Rust library}
\label{app:rust}

The solver provides both a direct API and a fluent builder:

\begin{lstlisting}[language={},%
  morekeywords={use,let,mut}]
use spectroxide::prelude::*;

let mut solver = ThermalizationSolver::builder(
    Cosmology::default()
)
    .grid(GridConfig::default())
    .injection(InjectionScenario::SingleBurst {
        z_h: 1e5,
        delta_rho_over_rho: 1e-4,
        sigma_z: 1000.0,
    })
    .build()
    .unwrap();

let result = solver.run_to_result(1e3);
\end{lstlisting}

Custom injection scenarios are supported via the \code{Custom} variant (see Sec.~\ref{sec:custom_scenarios} for examples in both Rust and Python).

\subsection{Python interface}
\label{app:python}

The Python package (\code{pip install -e ".[plot]"} from the \code{python/} directory) wraps the Rust PDE binary and provides a pure-Python analytic Green's function.
The unified entry point is \code{solve()}, which returns a \code{SolverResult} with attributes \code{x}, \code{delta\_n}, \code{mu}, \code{y}, \code{delta\_rho\_over\_rho}, and a \code{delta\_I} property for intensity in Jy/sr:

\begin{lstlisting}
import spectroxide

# --- PDE solver (full numerical solution) ---
result = spectroxide.solve(
    injection={"type": "single_burst", "z_h": 2e5},
    delta_rho=1e-5,
    method="pde",
)
print(result.mu, result.y)  # distortion parameters

# Dark photon depletion
result = spectroxide.solve(
    injection={"type": "dark_photon_resonance",
               "epsilon": 1e-6, "m_ev": 1e-5},
    method="pde",
)

# --- Green's function (fast, analytic) ---
result = spectroxide.solve(
    z_h=2e5, delta_rho=1e-5,
    method="greens_function",
)

# Custom heating rate via Green's function
import numpy as np
def dq_dz(z):
    """DM annihilation-like: dq/dz ~ (1+z)^2."""
    return 1e-25 * (1 + z)**2  # total drho/rho ~ 1e-6
result = spectroxide.solve(
    dq_dz=dq_dz, method="greens_function"
)
\end{lstlisting}

The same \code{dq\_dz} callable can be routed through the PDE solver by setting \code{method="pde"}, in which case the function is tabulated on a redshift grid and passed to the Rust binary automatically.
Custom cosmological parameters can be specified via the \code{Cosmology} class:
\begin{lstlisting}
cosmo = spectroxide.Cosmology.planck2018()
result = spectroxide.solve(
    injection={"type": "single_burst", "z_h": 1e5},
    delta_rho=1e-5,
    cosmo=cosmo, method="pde",
)
\end{lstlisting}

For multi-redshift sweeps (e.g., building a Green's function table), \code{run\_sweep()} parallelizes across injection redshifts using all available cores:
\begin{lstlisting}
sweep = spectroxide.run_sweep(
    delta_rho=1e-5,
    z_injections=[5e3, 2e4, 1e5, 5e5]
)
for r in sweep["results"]:
    print(r["z_h"], r["pde_mu"], r["pde_y"])
\end{lstlisting}

The package also exposes individual physics building blocks as standalone NumPy-vectorized functions, all importable directly from the top-level \code{spectroxide} namespace:
\begin{itemize}
    \item visibility functions: \code{j\_bb\_star}, \code{j\_mu}, \code{j\_y};
    \item spectral shapes: \code{mu\_shape}, \code{y\_shape}, \code{g\_bb};
    \item cosmological background quantities: \code{hubble}, \code{cosmic\_time}, \code{ionization\_fraction};
    \item distortion decomposition: \code{decompose\_distortion};
    \item unit conversion: \code{delta\_n\_to\_delta\_I}.
\end{itemize}

\paragraph{Tutorial notebooks.}
\label{app:notebooks}
The repository includes Jupyter notebooks that reproduce the figures in this paper and serve as starting points for new analyses:
\begin{itemize}
    \item \code{01\_getting\_started} --- Installation, first PDE run, and comparison of PDE vs.\ Green's function output.
    \item \code{02\_energy\_injection} --- Continuous injection scenarios: decaying particles, and $s$-wave versus $p$-wave DM annihilation.
    \item \code{03\_new\_physics} --- Dark photon oscillations and monochromatic photon injection via the PDE solver.
    \item \code{04\_custom\_scenarios} --- User-defined heating rates (\code{dq\_dz}) and tabulated photon sources.
    \item \code{05\_observational\_constraints} --- FIRAS limits on monochromatic photon injection and on dark-photon kinetic mixing.
    \item \code{06\_greens\_table} --- Building, caching, and convolving precomputed Green's function tables for fast parameter scans.
\end{itemize}
Additional notebooks under \code{notebooks/physics/} provide detailed validation: adiabatic-cooling checks, photon-injection benchmarks against the literature, and the resolution of the temporal injection width. Notebooks under \code{notebooks/observational/} derive FIRAS limits on photon injection and compare dark-photon limit conventions.
Self-contained notebooks that reproduce each paper figure are collected in \code{notebooks/paper\_figures/}. 

\end{appendix}

\end{document}